\documentclass[trackchanges]{aastex62}
\usepackage{graphicx}
\usepackage{float}
\usepackage{footnote}

\begin{document}

\def\ergcm{\hbox{erg cm$^{-2}$ s$^{-1}$ }}
\def\ergcmn{\hbox{erg cm$^{-2}$ s$^{-1}$}}

\author{J. Newman}
\affiliation{Department of Physics, Montana State University, Bozeman, MT 59717, USA}

\author{S. Tsuruta}
\affiliation{Department of Physics, Montana State University, Bozeman, MT 59717, USA}
\affiliation{Kavli Institute for the Physics and Mathematics of the Universe (IPMU) and World Premier Research Center Initiative (WPI), Tokyo University, Kashiwa, Chiba 277-8583, Japan}

\author{A. C. Liebmann}
\affiliation{Department of Physics, Montana State University, Bozeman, MT 59717, USA}

\author{H. Kunieda}
\affiliation{Department of Physics, Nagoya University, Furo-co, Chikusa-Ku, Nagoya 454-8602, Japan}

\author{Y. Haba}
\affiliation{Aichi University of Education, Kaiya, Aichi 448-843, Japan}

\title{Combined Analysis of X-Ray Spectra of NGC 3227}

\begin {abstract}

The 1.5 Seyfert galaxy NGC 3227 has been observed
by several X-ray missions. We carried out combined analysis of the data 
obtained by more recent major observations of this source - two observations performed
by \textit{XMM-Newton} in 2000 and 2006 and six observations performed
by \textit{Suzaku} in 2008. 

A unified model was constructed which is consistent with
all eight of the observations by the two satillites with large intensity and spectral changes. The model consists of a hard power law 
with the spectral index of $\Gamma_{Hard}=\replaced{1.27-1.6}{1.4-1.7}$ which is interpreted as the
Comptonized emission from the corona above an accretion disk.
In the high flux states an additional soft excess component dominates, which is consistent with a model with either a steeper power law with $\Gamma_{Soft}=\replaced{3.5-3.7}{3.3-3.85}$ or the warm Comptonization component. These emissions from the central engine are absorbed by a neutral partial covering material and warm absorbers.  A reflection component and several emission lines are also present. We examined the relationship between the intrinsic luminosity and the absorbers' physical parameters such as the column density, which suggests that the source expanded significantly during the bright states where the soft excess is greatly enhanced.   
\end{abstract}

\keywords{accretion, accretion disks - galaxies: active - galaxies: individual (NGC 3227) - galaxies: nuclei - galaxies: Seyfert - X-rays: galaxies}

\section{Introduction}

 The analysis and interpretation of NGC 3227 observations have been performed by various
authors throughout the years. Earlier X-ray observations from the \textit{Advanced Satellite for Cosmology and Astrophysics (ASCA)} taken in 1993 and 1995 and those by \textit{R\"{o}ntgen Satellite (ROSAT)} in 1993 gave evidence supporting the presence of both warm and neutral absorbers (Netzer et al. 1994, Ptak et al. 1994, Komossa \& Fink 1997, George et al. 1998). In the optical range the H$\alpha$/H$\beta$ ratio for both broad and narrow lines shows a degree of reddening consistent with the presence of dust (Komossa 2002, Cohen 1983, Gonzalez Delgado \& Perez 1997, Mundell et al. 1995, Rubin \& Ford 1968, Winge et al. 1995, Shull \& van Steenburg 1985).   Komossa \& Fink (1997) suggested the dust was part of the warm absorber similar to the concept of a ``dusty warm absorber" of IRAS 13349+2438 presented by Brandt, Fabian, \& Pounds (1996).  Kraemer et al. (2000) proposed that the warm absorber was too highly ionized to contain the dust. These authors suggested instead the presence of a second warm absorber at low ionization. The presence of this additional absorber was later supported by a \textit{Hubble Space Telescope (HST)}-STIS observation which detected intermediately ionized C, N, and Si in the optical/UV band (Crenshaw et al. 2001).

More recently \textit{XMM-Newton} observed NGC 3227 for 40 ks in 2000 and 108 ks in 2006. Subsequently \textit{Suzaku} observed this source six times with about a week between observations in 2008.
Gondoin et al. (2003) gave a model for the 2000 \textit{XMM-Newton} observation with power law continuum emission absorbed by a fully covering neutral absorber, a partially covering neutral absorber, and a fully covering warm absorber. A Gaussian
emission line to model the Fe K$\alpha$ line and an absorption edge
around 7.6 keV for the Fe absorption edge were also added.
Markowitz et al. (2009) presented a model for the 2006 \textit{XMM-Newton} observation of this source. 
It consists of a flatter primary hard power law emission with neutral absorption and an additional steeper power law soft excess.  Two zones of fully covering warm absorbers were applied to both. They also included Fe K$\alpha$ emission and an Fe absorption edge, as well as several emission lines. 
Noda et al. (2014) proposed a model for the six Suzaku observations. In their model the primary continuum emission consists of two power law components with different slopes; one steep with $\Gamma \sim$ 2.3 and another flatter with $\Gamma \sim$ 1.6 in the 2 - 50 keV band with no soft excess. The flat power law component is more absorbed while the steeper component is less absorbed. Among the six observations, the first is in a bright state while the rest are in the dim state. The flatter power law component appears in both dim and bright states while the steep power law appears and dominates in the bright state. These authors interpret this behavior as a phase transition from a dim state to a bright state which involves a change from flatter to steeper power law due to increased accretion rates, analogous to the stellar mass black hole case.   

The models presented for the two \textit{XMM-Newton} observations are somewhat different but they are still consistent with each other, noting that the source was in significantly different states. During the 2000 \textit{XMM-Newton} observation the source was in a substantially dim state and it is heavy absorbed by dense material in the line of sight. However, during the 2006 \textit{XMM-Newton} observation the source was in a brighter phase with much less absorption.

On the other hand we note that the Markowitz et al. (2009) model for the 2006 \textit{XMM-Newton} observation
and the Noda et al. (2014) model for the first of Suzaku observations, although both cases are during a similar bright phase, are very different.
For instance, in the bright state the primary hard power law continuum for \textit{XMM-Newton} (2006) by Markowitz et al. (2009) is flatter while for Suzaku by Noda et al. (2014) it is steep. Moreover the model by Markowitz et al. (2009) has additional soft excess while it is missing in the Noda et al. (2014) case. Two models for similar bright states of the same source are not consistent with each other. In other words, the models proposed for the individual observations, although they may be relevant by themselves, sometimes are not consistent with other observations.
Therefore, it will be worthwhile to carry out detailed combined analysis of all observations, two from the XMM-Newton and six from Suzaku, to explore a model for NGC 3227 which will be consistent with multiple observations from different missions combined.  That is the main goal of our current investigation.

NGC 3227 is an active galactic nucleus situated at RA 10h 23m 30.58s, Dec +10d 51m 4.18s (Anderson \& Ulvestad 2005).  The galaxy has a redshift of z = 0.00386 and central black hole mass of $M_{bh} = 4.22 \times 10^{7}$M$_\odot$ (Peterson et al. 2004). 

Section 2 explains the data reduction, Section 3 gives timing analysis, and
Section 4 is devoted to detailed spectral analysis. Section 5 gives discussions where a unified model is constructed based on the analysis and its comparison with other major earlier proposed models. Summary and concluding remarks are given in Section 6.

\section{Observations and Data Reduction}

Two \textit{XMM-Newton} observations were taken six years apart in 2000 and 2006. Eight \textit{Suzaku} observations were taken
in 2008 with about one week in between each observation. Summary of observation start times, end times, and usable exposure
times are given in Table~\ref{tab:xob} for both \textit{Suzaku} and \textit{XMM-Newton}. 
 
The range of 0.3-10 keV was used for the European Photon Imaging Camera (EPIC) pn (Str\"uder et al. 2001) for the \textit{XMM-Newton} observations.  The Metal Oxide Semi-conductor (MOS) (Turner et al. 2001) cameras were not used as this paper focuses on the pn camera.  For \textit{Suzaku} the 0.5-10 keV range was used for the X-ray Imaging Spectrometer (XIS) cameras (Koyama et al. 2007) and the 15-50 keV range for
the Hard X-ray Detector (HXD)-PIN.  \added{\bf{For the pn and XIS cameras the 1.7-2.1 keV range was omitted due to calibration uncertainties.}}  All spectra were analyzed using XSPEC version 12.9.0i (Arnaud 1996).

All Observational Data Files (ODF) were downloaded using the w3browse HEASARC tool
from the NASA website.  The observations are numbered in sequential order based on the date of observation.
Furthermore, an ``X'' prefix is used if observed by \textit{XMM-Newton} or an ``S'' prefix if performed
by \textit{Suzaku} (e.g., X1, S3 etc.). \textit{XMM-Newton} ODF data are processed using the xmmextractor command as included
in Scientific Analysis System (SAS) version 16.0.0. \ \textit{Suzaku}
data were processed using XSELECT version 2.4c. Non X-ray background files were
taken from the HXD-PIN tuned non X-ray background database from the
NASA website.   All \textit{Suzaku} extraction regions were a circle of 130'' for the source and an annulus of inner radius 200'' and outer radius of 300'' for the background.  The first \textit{XMM-Newton} observation (X1) used a circular region of 40" for the source and an annulus of inner radius of 60" and outer radius of 82.46" for the background, while the second observation (X2) used a source circle of 40" and a background annulus with inner and outer radii of 60" and 96.01" respectively.  

In order to simultaneously use the XIS and HXD-PIN data a calibration constant is needed.  This constant is obtained by using the method described in Walton et al. (2013) and the X-ray Telescope (XRT) response given by Maeda et al. (2008). The XIS-PIN calibration
constant is a multiple of the XIS1-XIS03 calibration constant and
1.16. The XIS1-XIS03 calibration constant is found by fitting a galactic
absorbed broken power law to each Suzaku observation. The regions 1.7-2.4 keV
and 4-7 keV are ignored to remove calibration uncertainties and Fe contamination
respectively. A constant is multiplied by the model. All parameters
between XIS 0 \& 3 and XIS 1 are tied with the exception of the constant.  After
fitting, the value of the XIS 1 constant is the XIS1-XIS03 calibration
constant. 

Our models adopt $H_{o} = 70, q_{o} = 0.0$, and $\Lambda_{o} = 0.73$.  For the \textit{Suzaku} observations the
XIS 0 and XIS 3 data sets are combined as both are front illuminated. In
each model all components are attenuated with a neutral hydrogen column
density of $N_H = 1.99 \times 10^{20}cm^{-2}$ reported by the Galactic Column Density
HEASARC tool (Angelina \& Sabol), with data taken from Kalberla et al. 2005.  All fit parameters are given in the source rest frame and errors are reported at the 90$\%$ confidence level ($\Delta\chi^2  = 2.7$) unless otherwise stated.

\begin{table}[htb!]
\resizebox{0.8\textwidth}{!}{\begin{minipage}{\textwidth}
\caption{NGC 3227 Observation Summary: Start time (UTC), end time (UTC) and exposure times (ks) are given. For the exposure times XIS and HXD-PIN of \textit{Suzaku} while the pn for \textit{XMM-Newton} was used.}
\begin{tabular}{l l l l  c c} 
\hline
Observation & Year & Start Time (UTC) & End Time (UTC) & XIS/pn Exposure Times (ks)  & HXD Exposure Times (ks) \\
\hline \hline
703022010 (S1) & 2008 & Oct. 28 08:12:52 & Oct. 29 00:34:49 & 58.92 & 48.07 \\
703022020 (S2) & 2008 & Nov. 4 03:36:31 & Nov. 4 18:31:01 & 53.7 & 46.74 \\
703022030 (S3) & 2008 & Nov. 12 02:48:55 & Nov 12. 18:31:47 & 56.57 & 46.68 \\
703022040 (S4) & 2008 & Nov. 20 17:00:00 & Nov. 21 10:56:08 & 64.57 & 43.43 \\
703022050 (S5) & 2008 & Nov. 27 21:29:20 & Nov. 28 19:33:11 & 79.43 & 37.42 \\
703022060 (S6) & 2008 & Dec. 2 14:28:03 & Dec. 3 04:44:04 & 51.41 & 36.91 \\
0101040301 (X1) & 2000 & Nov. 28 18:15:41 & Nov. 29 05:26:33 & 27.28 &  \\
0400270101 (X2) & 2006 & Dec. 3 01:53:31 & Dec. 4 08:09:51 & 89.77 &  \\
\hline 
\end{tabular} 
\end{minipage}}
\label{tab:xob}
\end{table}

\section{Timing Analysis}

\subsection{\textit{Light Curves}}

\begin{figure}[htb!]
\includegraphics[scale=0.6]{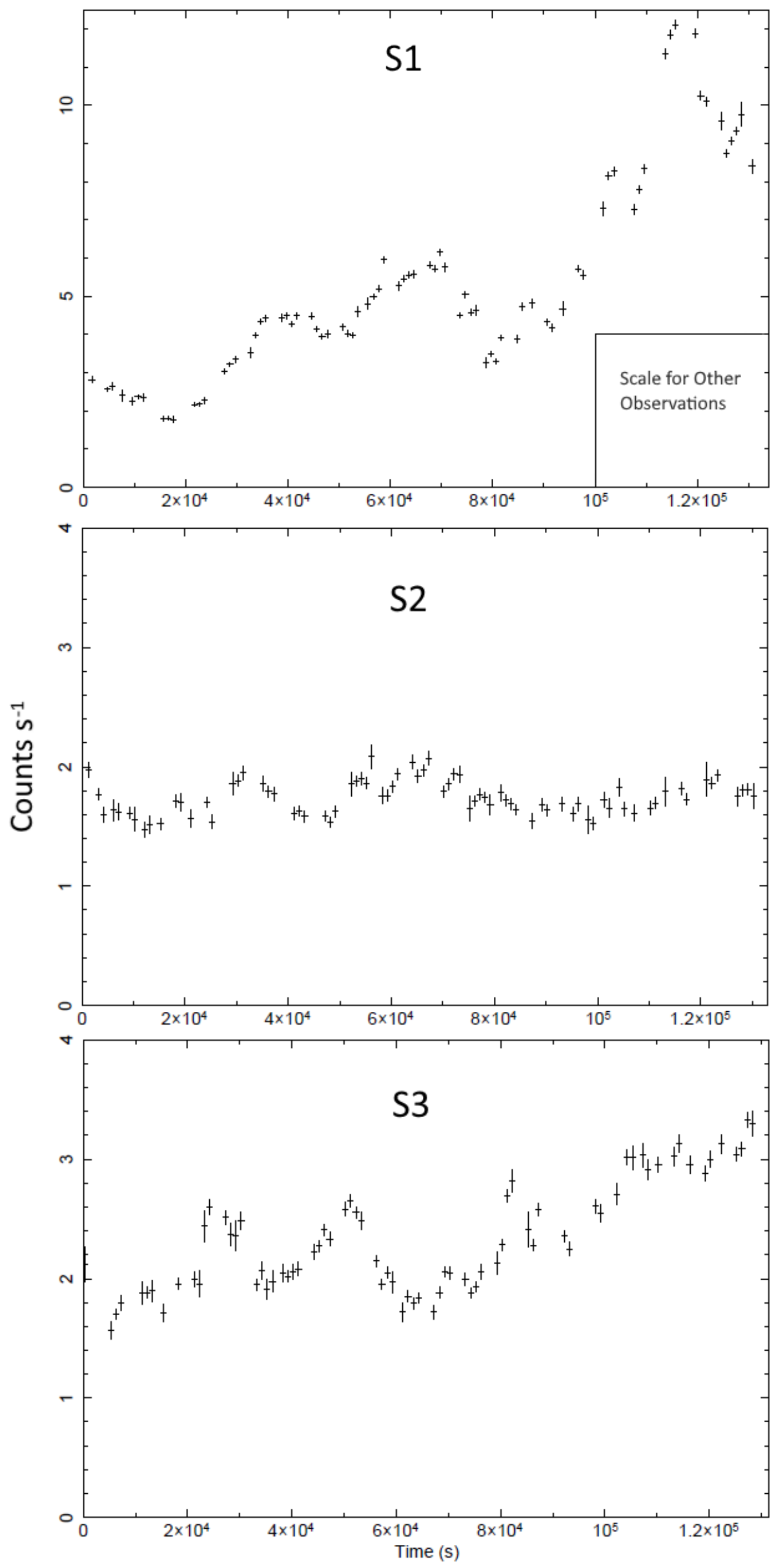}\includegraphics[scale=0.6]{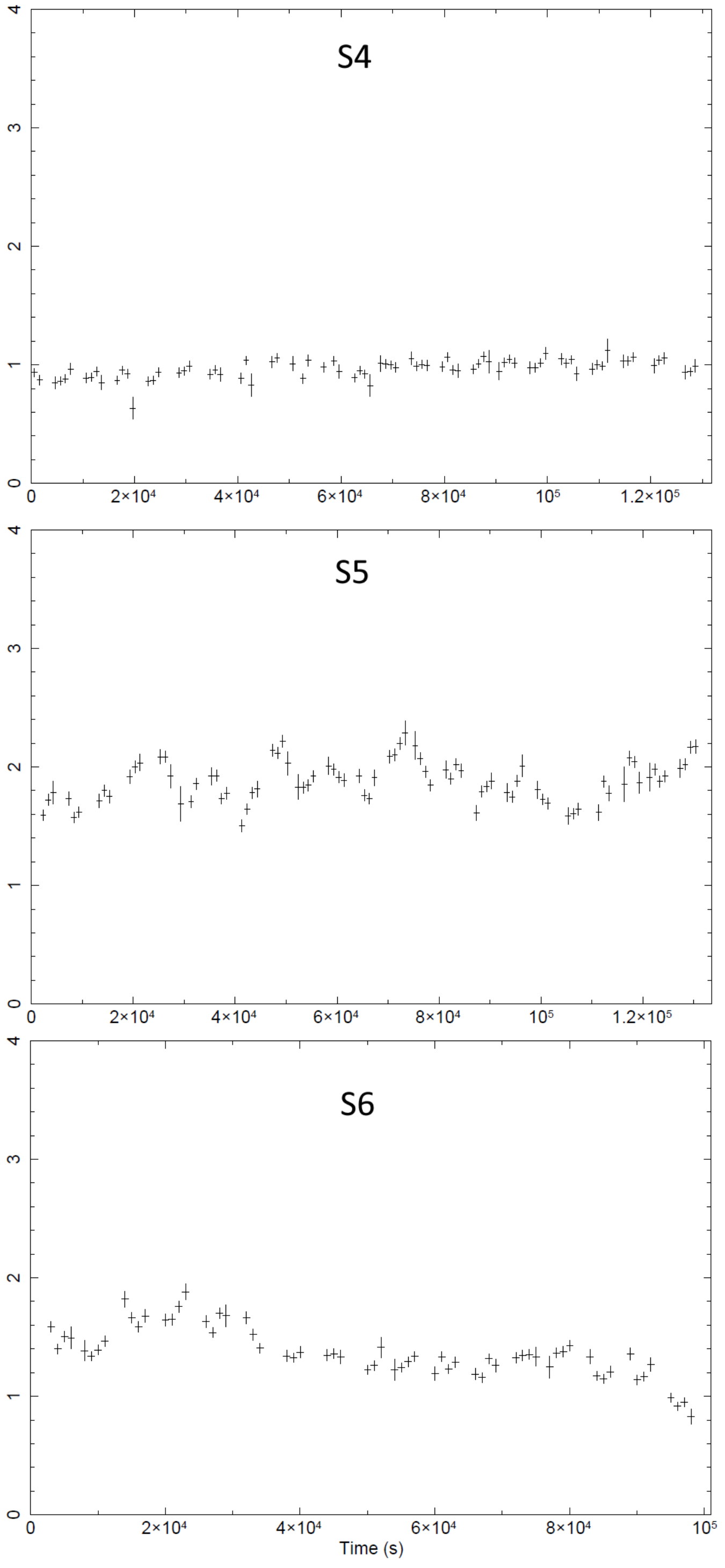}
\caption{The light curves are shown for six \textit{Suzaku} observations S1 to S6. The energy range used is 0.5 to 10 keV.
Since the intensity of S1 is high compared with other observations the vertical axis is expanded.  The insert corresponds to the vertical scale of other observations.}
\label{fig:slc}
\end{figure}

\begin{figure}[htb!]
\includegraphics[scale=0.6]{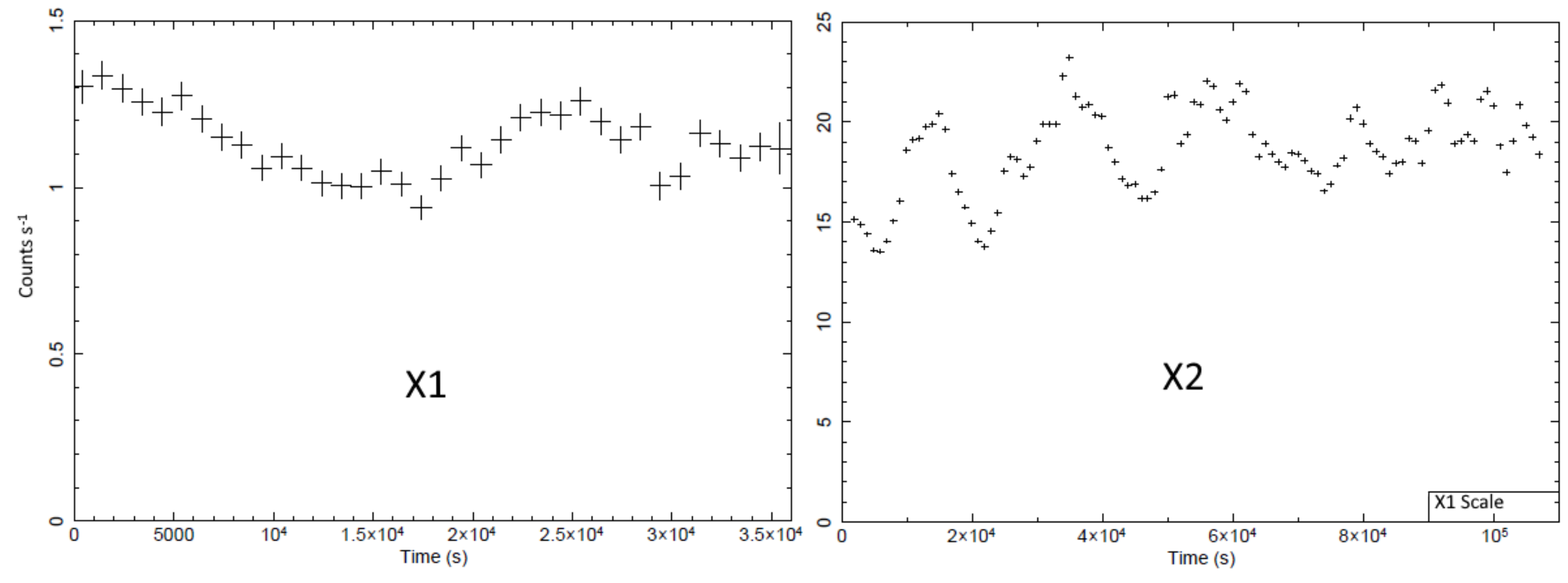}\protect
\caption{\textit{XMM-Newton} Light curves are shown for observations X1 and X2. The energy range used is 0.3 to 10 keV. We note that X1 was in a dim state while X2 was in a bright phase.  The insert in the X2 plot is the vertical scale of the X1 plot.}
\label{fig:xlc}
\end{figure}

To begin with we extracted the light curves of all observations. Figure~\ref{fig:slc} 
shows the light curves for the \textit{Suzaku} observations
using all 3 XIS cameras. Figure~\ref{fig:xlc} 
shows the \textit{XMM-Newton} light curves
using the pn camera. The \textit{Suzaku} light curves used 0.5-10 keV while the \textit{XMM-Newton} light curves used 0.3-10 keV. 
Both \textit{Suzaku} and \textit{XMM-Newton} light curves exhibit similar
characteristics. The variability increases with flux intensity.
In the lowest flux states (S4 and X1) light curves are nearly flat with little variability.
In low but slightly brighter states (S2, S5, and S6) the fluxes show some variability.
The moderate flux state (S3) shows significant changes in
count rates. Both of the highest flux states S1 and X2 are highly
variable. Furthermore the later half of S1 exhibits higher flux and variability within the observation. The two highest states S1 and X2 may be variable enough to perform time-resolved analysis (see a subsequent paper).  Note that both \textit{XMM-Newton} observations exhibited a similar intensity of variability ($\sim30\%$) despite having significantly different fluxes.

\subsection{\textit{Flux-Flux Plot}}

\begin{figure}[htb!]
\includegraphics[scale=1]{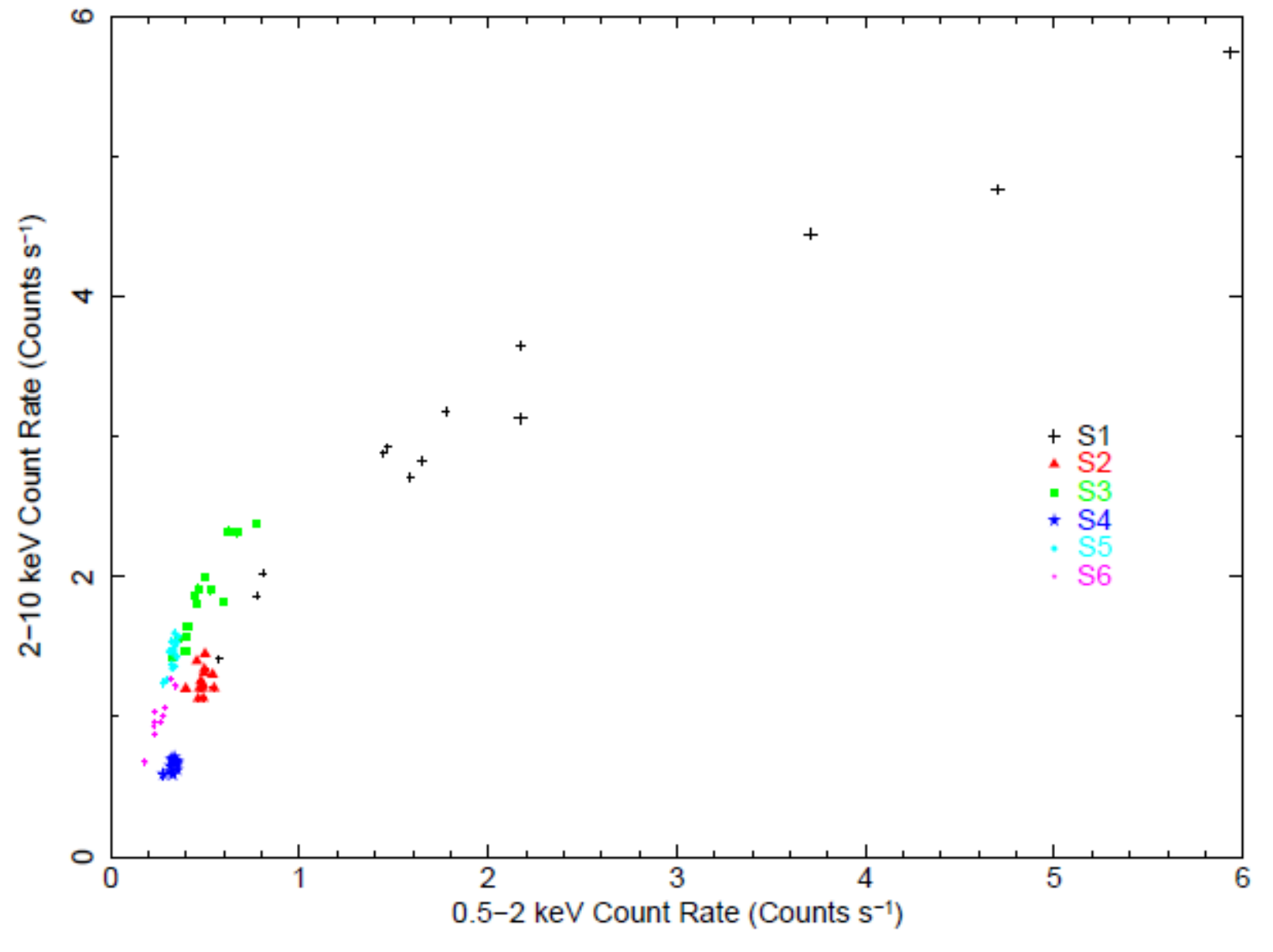}
\caption{The 2-10 keV vs 0.5-2 keV flux-flux plot is shown for all Suzaku observations S1 to S6. Since the slope from the origin to the data point tells the hardness ratio, the brighter phase S1 has a softer spectrum compared with other fainter phases with steeper slope.}
\label{fig:scc}
\end{figure}

\begin{figure}[htb!]
\includegraphics[scale=1]{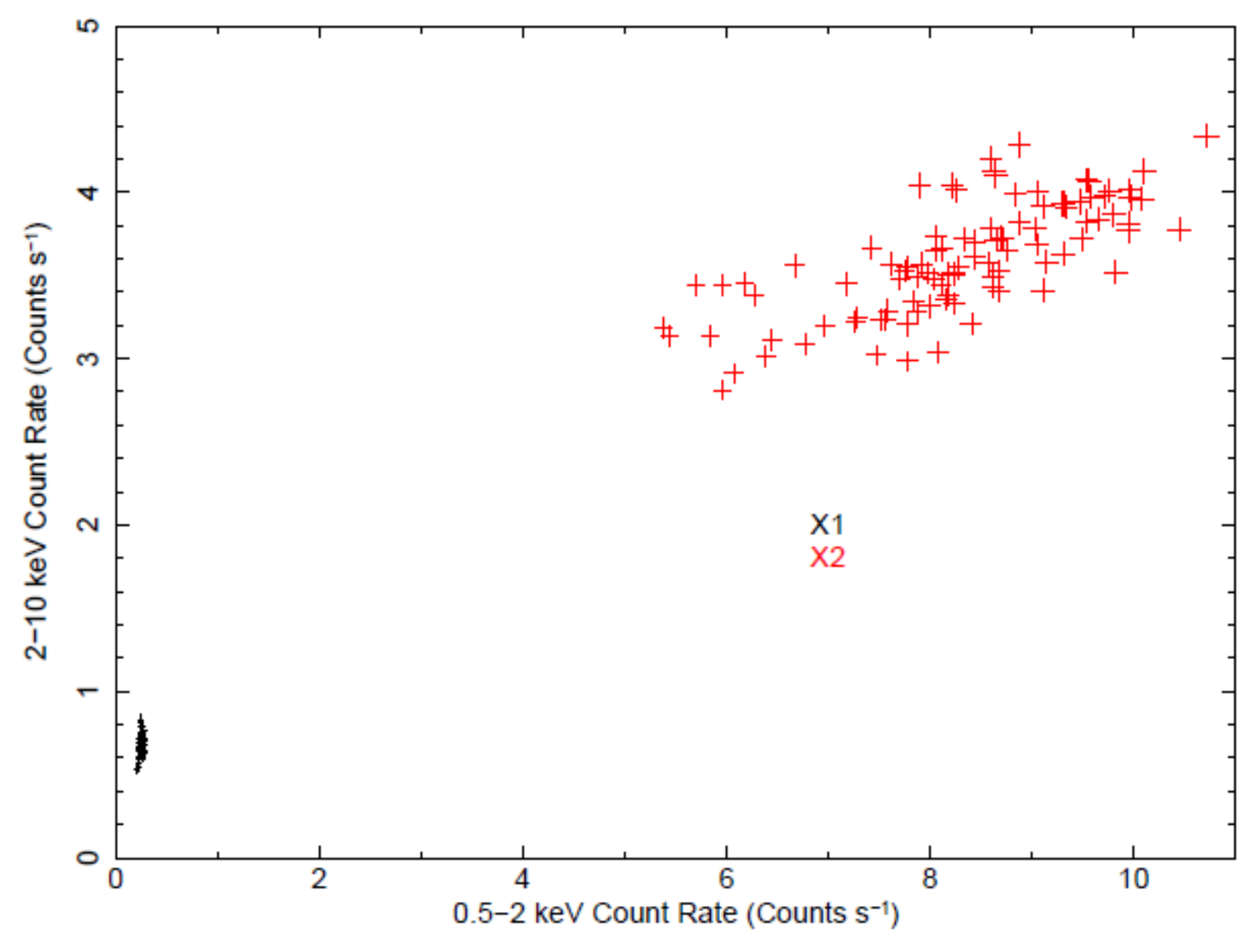}
\caption{The 2-10 keV vs 0.5-2 keV flux-flux plot for two XMM-Newton observations X1 and X2. }
\label{fig:xcc}
\end{figure}

After extracting light curves of the data we formed flux-flux plots
for the source, in order to examine spectral characteristics of the soft and hard bands. The 0.5-2 keV range is used for the soft band and 2-10 keV for the hard band. The vertical and horizontal axes are flux of the hard and soft bands respectively.  
The flux-flux plot for \textit{Suzaku} was formed by summing data from all
three XIS detectors while the flux-flux plot for \textit{XMM-Newton} was created using the pn camera.  The flux-flux plots are displayed
in Figures~\ref{fig:scc} and~\ref{fig:xcc}
 for \textit{Suzaku} and \textit{XMM-Newton}, respectively.

The flux-flux plot for \textit{Suzaku} shows two distinct branches.
The left branch (dim branch) has a steeper slope and shows the source
in a low flux state. In the right branch (bright branch) the count
rates of the soft band start to show a larger increase than the hard band.
The transition between the two branches occurs between 0.6-1.8 $counts \ s^{-1}$ for the soft band and
2.2-3.1 $counts \ s^{-1}$ for the hard band. The only two observations that
encounter this transition are S1 and S3. With further increase
in intensity the source enters the bright branch which
corresponds to a high flux state. This branch has a flatter slope than
the dim branch. 

With only two observations \textit{XMM-Newton} was unable to observe
the transition phase. However, it was able to see the source in its dim state
during X1 and in its bright state during X2. The results are consistent with the
\textit{Suzaku} flux-flux plots. The dim state showed a steeper slope
while the bright state had a more gradual trend. 

The property and implication of these distinct two branches will be further examined and
discussed in Sections 4 and 5. Note that both of the observations that showed
high temporal variability, S1 and X2, spent most or all of
their time in the bright state. The slope flattening in the bright branch suggests an excess of emission in the soft band.  This soft excess emission is highly variable. This highly variable soft excess emission will be examined further in a subsequent paper where the time-resolved analysis will be applied.  

\section{Spectral Analysis}

\subsection{\textit{Suzaku Spectral Analysis}}

Spectral analysis was performed with the six \textit{Suzaku} observation data observed by the three functional XIS cameras as well as the HXD-PIN. The \replaced{2}{2.1}-10 keV range was considered first with the XIS and a model fit of that range was made.  Then the 15-50 keV HXD-PIN data was added. Once an acceptable fit for the \replaced{2}{2.1}-50 keV range was achieved, this model was extended down to 0.5 keV to form our final best fit broadband model.

\subsubsection{\textit{Suzaku 2.1-50 keV Band Analysis}}

\begin{figure}[htb!]
\includegraphics[scale=1]{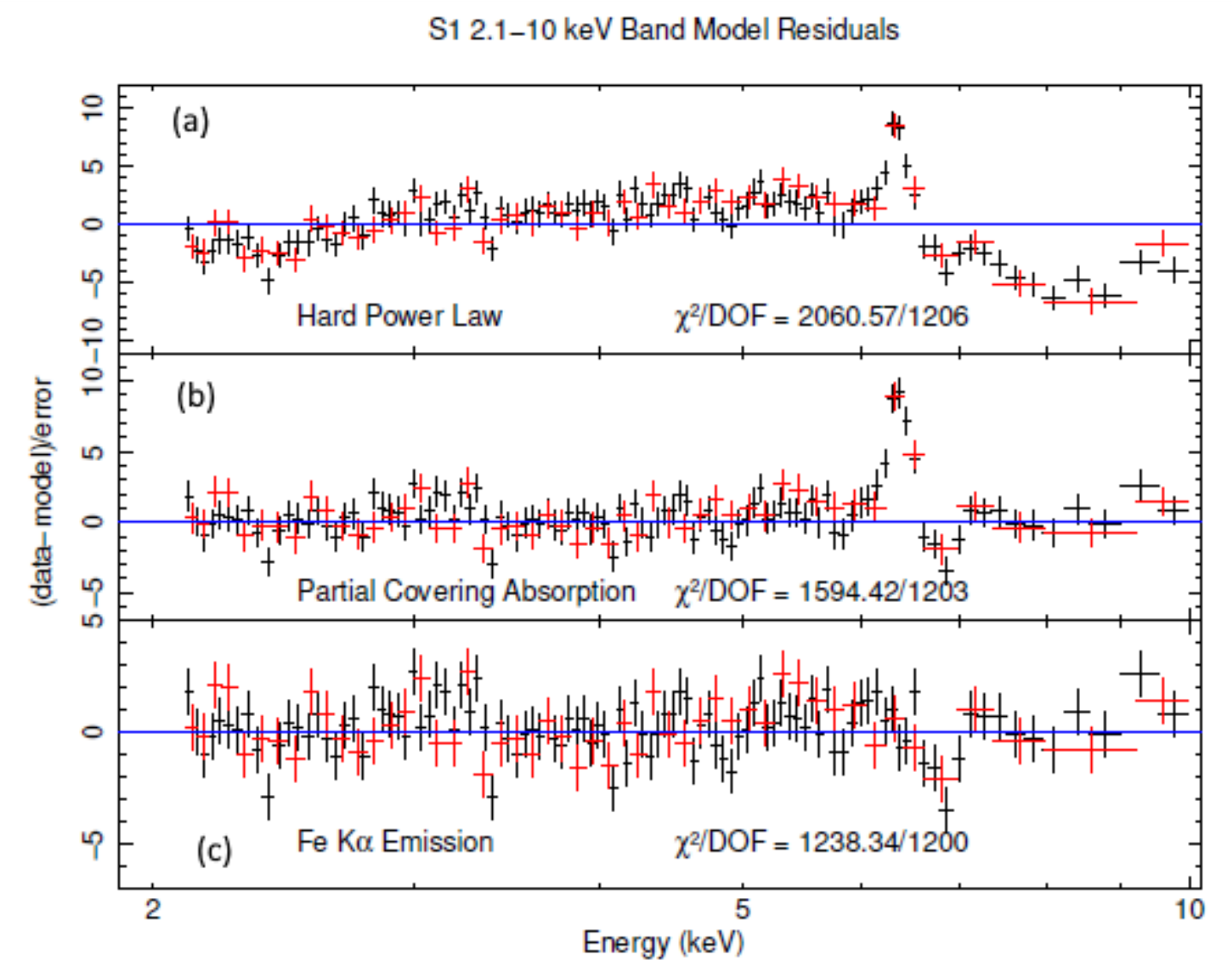}
\caption{S1 Hard X-ray Band \replaced{(2-10 keV)}{(2.1-10 keV)} Spectral Fits. The ratio of the data to the model is shown. XIS 0 \& 3 are the black data and XIS 1 is red. At first a simple power law was applied with the galactic absorption (Figure 5a).  Then a \replaced{neutral}{\bf{ionized partial}} covering was added
(Figure 5b). A gaussian at 6.4 keV was introduced to eliminate the residual due to the Fe K$\alpha$ line (Figure 5c).} 
\label{fig:s1med}
\end{figure}

The spectral fits were started with the S1 data between \replaced{2}{2.1} and 10 keV and the residual of the data from the model for S1 is plotted in Figure~\ref{fig:s1med}.
Assumed model is the simple redshifted power law (model ``zpowerlw'')
with galactic absorption (model ``phabs'') to the
2.1-10 keV spectrum. This showed significant residuals around 6.4 keV and below 3 keV (see Figure~\ref{fig:s1med}a). 
Note that all redshifted model components are set to the redshift (z = 0.00386) of NGC 3227. Adding the \replaced{neutral}{\bf{ionized}} partial covering model (``\replaced{zpcfabs}{\bf{zxipcf}}'') eliminated most of the residuals below 3 keV (see Figure~\ref{fig:s1med}b). 
Next the 6.4 keV residuals were modeled with a redshifted gaussian model (``zgauss''), assuming
Fe K$\alpha$ emission, to achieve the best model in the \replaced{2}{2.1}-10 keV band (see Figure~\ref{fig:s1med}c). 

\begin{figure}[htb!]
\includegraphics[scale=1]{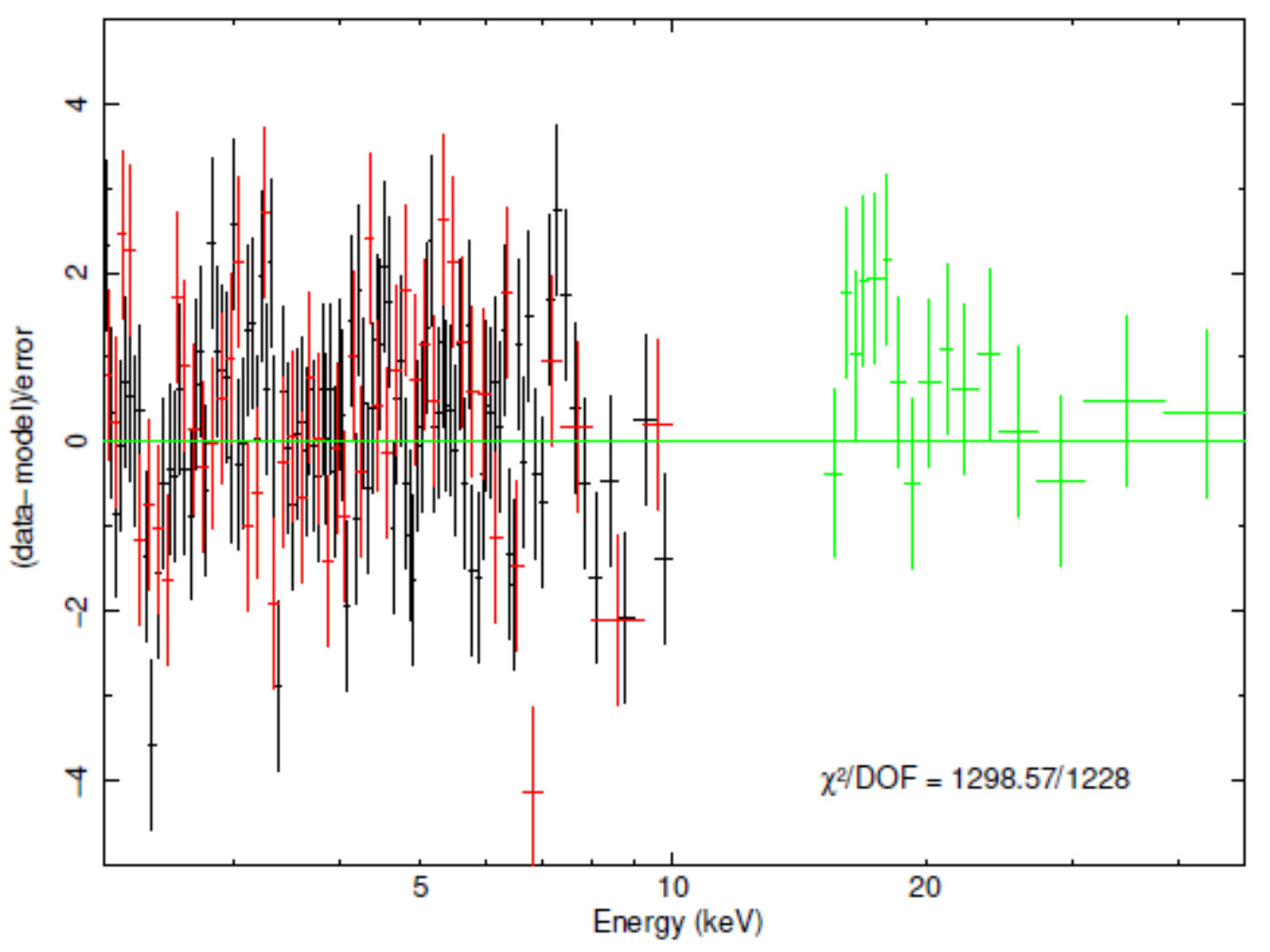}
\caption{S1 broader Hard X-ray Band (\replaced{2}{2.1}-50 keV) Spectral Fit. The ratio of the data to the model is shown. XIS 0 \& 3 are the black data, XIS 1 is red, and HXD-PIN is green. After 15-50 keV HXD-PIN data were added to the \replaced{2}{2.1}-10 keV model, a reflection model was applied to obtain the best fit \replaced{2}{2.1}-50 keV model.} 
\label{fig:s1ref}
\end{figure}

Afterwards the 15-50 keV HXD-PIN data are included with the appropriate calibration constant (model ``constant''). 
Since the residuals to the HXD data suggested a reflection component, an accretion disk reflection model (model ``xillver" Garcia et al. 2013) was added with inclination angle set to 60 degrees and solar abundance (see Figure~\ref{fig:s1ref}). \added{\bf{In this model the AGN accretion disk is responsible for the reflection component of the emission.}}  In addition an Fe K edge at 7.11 keV was added.  \added{\bf{There existed a narrow absorption residual at $\sim$7 keV which was modeled by an inverse guassian for $\Delta$$\chi^2 = 25.14$.  This is most likely systemic as it was not seen in most other observations.  A similar residual was found in S5 but was not as strong.  It is most likely due to Xillver overestimating Fe contribution at this energy.}}  In this way the best fit \replaced{2}{2.1}-50 keV model was acquired.  \deleted{Residuals remain above 6 keV for the XIS cameras however these are reduced when additional absorption is added during broad band model formation.}

The same procedure was applied to all data from the six \textit{Suzkau} observations and similar results were obtained.  \deleted{The exception was the lowest flux \textit{Suzaku} state (S4).  This observation shows such strong absorption below 6 keV that a single partial covering model was insufficient.  This issue is alleviated during the broad band analysis (see Section 4.1.2).}  \replaced{Furthermore}{\bf{In S4}} the Fe K$\alpha$ line was broader than the xillver parameters anticipated.  An additional guassian at 6.4 keV fit the residual with $\Delta$$\chi^2 = 30.24$.  \deleted{The observation S5 also contained a sharp absorption residual at 7 keV.  This is well fit by a narrow guassian ($\Delta$$\chi^2 =32.5$) and is most likely systemic as it did not appear elsewhere.}

\subsubsection{\textit{Suzaku Phenomenological Broad Band 0.5-50 keV Analysis}}

To model the full 0.5-50 keV broad band data, the \replaced{2}{2.1}-50 keV band model obtained in Section 4.1.1 was extended down to 0.5 keV. On the lower energy side the residuals showed a large trench centered near 0.9 keV consistent with warm absorption.  Other work such as Komossa \& Fink 1997 have also found evidence for warm absorption.  For our warm absorber we used an XSTAR table (XSTAR v2.39 Kallman 2019). For all XSTAR tables, an $\alpha$ parameter was used which is equivalent to the hard X-ray power law index.  \added{\bf{Most other parameters using the ``xstar2xspec" command were left as the default values including the velocity ($v_{turb} = 300$ $km$ $s^{-1}$).  The range of allowed column densities was expanded as needed.}}  Though one fully covering warm absorber was applied to the data, significant residuals still remained.  The second warm absorber at higher ionization was added.  The parameters of the high ionization warm absorber were less constrained \deleted{and only upper limits were set for S2}.  Although less significant than the low ionization warm absorber, this second warm absorber was necessary to properly model S1.  Other observations did not strictly require the high ionization warm absorber for an acceptable fit but were improved by including it.  The one exception was S4 which did not require a second warm absorber.  A single warm absorber was sufficient to form an acceptable fit and attempting to force a second warm absorber only worsened the fit.  \added{\bf{Henceforth all references to the high ionization warm absorber and low ionzation warm absorber refer to these two fully covering zones instead of the partially covering warm absorber.}}

In the brightest state (S1) this model \replaced{had the proper shape but
the soft 0.5-2 keV range was underestimated}{\bf{underestimated the 0.5-1.4 keV range}}. Therefore, a power law was added to model the extra soft emission.  This soft power law is much steeper ($\Gamma_{Soft} = \replaced{3.62_{-0.38}^{+0.42}}{3.31_{-0.11}^{+0.11}}$) than the hard power law contained within the xillver parameters.  In other observations the additional soft excess features were absent. 

With the warm absorbers added there now existed positive residuals
around 0.58 keV in \replaced{S4}{\bf{S3, S4, S5}} and S6, which
are consistent with an O VII line.  A gaussian emission feature was introduced with energy at 0.58 keV and narrow width ($\sigma$ = $10^{-5}$) to model this.  This improved the fit by \replaced{$\Delta$$\chi^2 = 11.06$ for S4 and $\Delta$$\chi^2 = 17.11$ for S6}{\bf{$\Delta$$\chi^2 = 15.86$ for S3, $\Delta$$\chi^2 = 4.18$ for S4, $\Delta$$\chi^2 = 7.6$ for S5, and $\Delta$$\chi^2 = 9.38$ for S6}}.
 Note that the O VII line was detected by XIS 1 only.  This is because XIS 1 is more sensitive to softer X-rays compared to XIS 0 \& 3.  All other emission lines were also modeled with thin redshifted guassians.  \deleted{There also existed residuals around 0.78 keV in S3.  This is consistent with an O VIII line.  Adding another \deleted{thin redshifted }gaussian at 0.775 keV modeled this emission and improved the fit by $\Delta$$\chi^2 = \replaced{24.32}{50.06}$.}  In S3, S4, and S5 there were residuals around 0.9 keV which is consistent with a Ne IX line.  Adding a guassian fixed at 0.922 keV improved the fit by \replaced{$\Delta$$\chi^2 = 23.22$ for S3, $\Delta$$\chi^2 = 13.46$ for S4, and $\Delta$$\chi^2 = 83.67$ for S5}{\bf{$\Delta$$\chi^2 = 12.83$ for S3, $\Delta$$\chi^2 = 11.92$ for S4, and $\Delta$$\chi^2 = 32.95$ for S5}}.  In S4 there existed emission residuals around 1.02 keV which are consistent with a Ne X line. This was modeled with a gaussian fixed at 1.022 keV, which improved the fit by $\Delta$$\chi^2 = \replaced{16.56}{10.38}$.  Note that this may be systemic as we did not detect this line in other observations.  However, Ne X absorption at this energy appeared in the Reflection Grating Spectrometer (RGS) analysis of X2 performed by Markowitz et al. 2009 (see their Section 5) suggesting presence of this Ne species.  Emission from O VII\deleted{, O VIII,} and Ne IX were detected during the Turner et al. 2018 RGS analysis of this source.  Markowitz et al. 2009 also detected absorption from those same three species.  \added{\bf{Ne IX and Ne X emission are often associated with starburst activity in galaxies.  Starburst activity was found by Rodriguez-Ardila \& Viegas 2003 as well as Davies et al. 2006 and evidenced by the detection of the 3.3 $\mu$m Polycyclic Aromatic Hydrocarbon feature by Imanishi 2002.  With these sources we consider starburst activity to be the source of this Ne emission in NGC 3227.  We note that S1 did not detect any of the lines.  This is most likely because the soft excess emission has outshined the lines and made them not visible.  Including these lines in S2 did not have a meaningful effect on the fit (combined Ne IX and O VII $\Delta$$\chi^2 = 0.3$) and were omitted from the model.  }}Emission line energies were verified using the online atomic database AtomDB version 3.0.9.  \added{\bf{As an alternative to using guassian emission lines we considered utilizing the ``mekal" model (Mewe et al. 1985, Mewe et al. 1986, Liedahl et al. 1995). While a plasma temperature of ~700-1000 eV gave respectable fits they were worse than using individual emission lines for the \textit{Suzaku} observations.  For X1 it signficantly worsened the fit with $\Delta$$\chi^2 = 48.03$.  For this reason we have chosen to use emission lines. }}

\begin{figure}[htb!]
\includegraphics[scale=0.75]{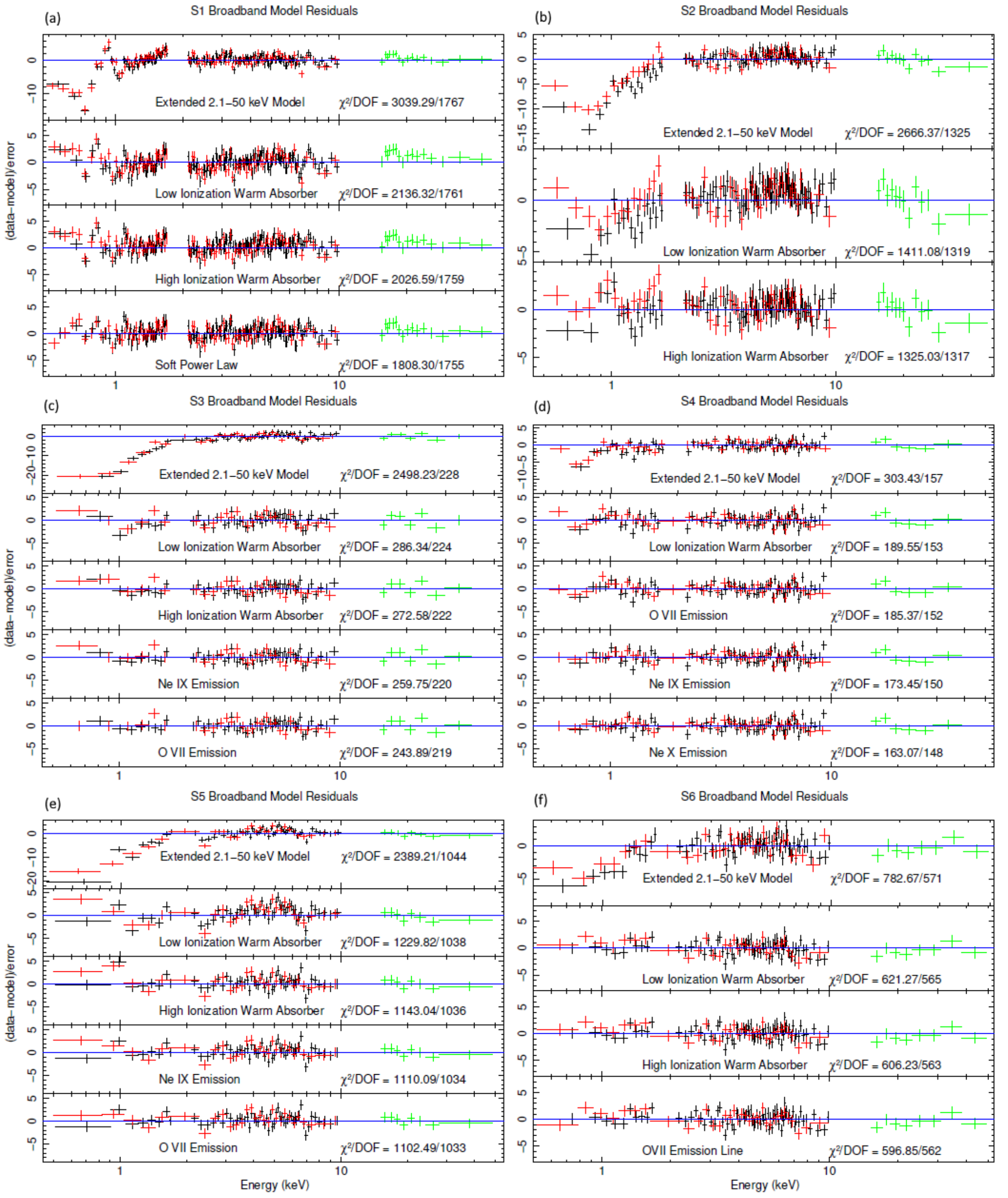}
\caption{S1 (a), S2 (b), S3 (c), S4 (d), S5 (e) and S6 (f) Spectral Fits. The ratio of the data to the model is shown. In S1 after the \replaced{2}{2.1}-50 keV model is extended to 0.5-50 keV and adding the low ionization and high ionization warm absorbers, the soft power law is applied to obtain the best fit. The warm absorbers are added to S2.  Emission lines are added to S3, S4, S5, and S6 data after the warm absorbers.  \replaced{An O VII line is added to S4 and S6, an O VIII line is added to S3, a Ne IX line is added to S3, S4, and S5, and a Ne X line is added to S4.}{An O VII line is added to S3, S4, S5, and S6.  A Ne IX line is added to S3, S4, and S5 and a Ne X line is added to S4}}
\label{fig:s1s2res}
\end{figure}

The 0.5 - 50 keV broad band spectral models thus obtained for observations S1 to S6 are displayed in Figure~\ref{fig:s1s2res}.
Model parameters are listed in Table~\ref{tab:spar}.

\deleted{Two characteristics are apparent. The first one is that the hard photon index tends
to increase with flux (see Table 2). The second one is that the highest flux state (S1) had the lowest covering fraction. The covering fractions of other observations were somewhat higher and similar. This issue will be discussed further in Section 4.4.}

\subsection{\textit{XMM-Newton Spectral Analysis}}

The EPIC pn spectra of two \textit{XMM-Newton} observations X1 and X2 were \deleted{simultaneously}analyzed by using SAS. 
\deleted{All parameters except normalization were tied between them.}  As with \textit{Suzaku} both cases included galactic hydrogen column density.  In both observations initially, an acceptable model was fitted to the \replaced{2}{2.1}-10 keV data, then it was extended to the band of 0.3-10 keV and modified to achieve a good fit.

Analysis of the RGS data from \textit{XMM-Newton}
 X2 observation has already been performed thoroughly by Markowitz et al. (2009) and as such we did not perform in depth spectral analysis using this instrument.  They found evidence supporting the existence of two separate warm absorbers.  For details see Markowitz et al. 2009 Section 5.  The X1 observation took place during a dim state and hence the data was not good enough for RGS analysis.  
		
\subsubsection{\textit{XMM-Newton 2.1-10 keV Band Analysis}}

First attempt for a single power law model fit to the X1 data did not provide a good fit with residuals
across the majority of the band, while the power law provided a decent fit to X2 except for residuals around the Fe line.  To keep consistancy with the \textit{Suzaku} observations the power law model was replaced with the xillver model, which includes the Fe K$\alpha$ line and reflection features.

At this point X1 still had significant residuals below 4 keV and required
some absorption feature. These residuals were reduced by \replaced{a}{\bf{an ionized}} partial covering model as was found in \textit{Suzaku} spectral analysis. 
On the other hand the X2 spectrum was reasonably well fitted with just the photo-absorbed xillver model.  Partial covering model required a covering fraction $>0.95$ so a fully covering neutral absorber model (``zphabs") was introduced instead.  \added{\bf{Using an ionized absorber instead worsened the fit.}}  \deleted{Consistent with \textit{Suzkau}, the high flux state has a higher hard photon index than a low flux state.   }

\subsubsection{\textit{XMM-Newton Phenomenological Broad Band 0.3-10 keV Analysis}}

After obtaining the \replaced{2}{2.1}-10 keV band best fit model it was extended to the broader
0.3-10 keV band.  For X1 the negative residuals in the soft band were consistent with the warm absorbers (Komossa \& Fink 1997). After applying the low ionization warm absorber emission residuals remained near 0.42 keV.  This energy is consistent with a N VI emission line.  An additional narrow gaussian fixed at 0.42 keV provided a good fit by $\Delta$$\chi^2 = \replaced{61.64}{94.94}$.  There remained residuals around 0.9 keV thus a 0.922 keV (Ne IX) guassian was added which improved the fit by $\Delta$$\chi^2 = \replaced{19.34}{14.97}$.

For observation X2 distinct positive residuals were found below \replaced{2}{1.7} keV, which suggested a second continuum power law to model this excess soft X-ray emission as was seen for S1.
Additional second power law of index $\Gamma_{Soft} = \replaced{3.72_{-0.09}^{+0.09}}{3.83_{-0.02}^{+0.02}}$ significantly improved
the fit.  Since the residuals seemed to be consistent with warm absorption\deleted{ feature}, two zones of warm absorber model were applied.  \replaced{Then}{\bf{With this change}} most of the residuals had dropped to an acceptable level. There remained emission residuals near
0.58 keV and absorption residuals around 0.74 keV. The 0.58 keV residual is
most likely O VII emission and was also detected by Markowitz et al. 2009. This was modeled with a guassian which improved the fit by $\Delta$$\chi^2 =\replaced{75.84}{62.2}$.  \added{\bf{Compared to \textit{Suzaku's} XIS cameras, the pn has better spectral resolution at this energy which is why it was able to detect this line while S1 could not.  Attempting to force this line into X1 significantly worsened the fit ($\Delta$$\chi^2 = -40.98$).}}  The 0.74 keV absorption feature was modeled with a redshifted absorption edge and is consistent with O VII edge absorption and possibly the Fe unresolved transition array (UTA) (Sako et al. 2001, Makowitz et al. 2009).  This is most likely part of the warm absorbers and its inclusion improved the fit by $\Delta$$\chi^2 = \replaced{482.05}{473.32}$. \deleted{There remained a distinct absorption feature just below 2 keV.  This is well modeled by an absorption edge centered at 1.84 keV (Si K edge) with $\Delta$$\chi^2 = \replaced{109.09}{23.53}$.  This feature is identified as instrumental by Markowitz et al. 2009.}

\begin{figure}[htb!]
\includegraphics[scale=0.6]{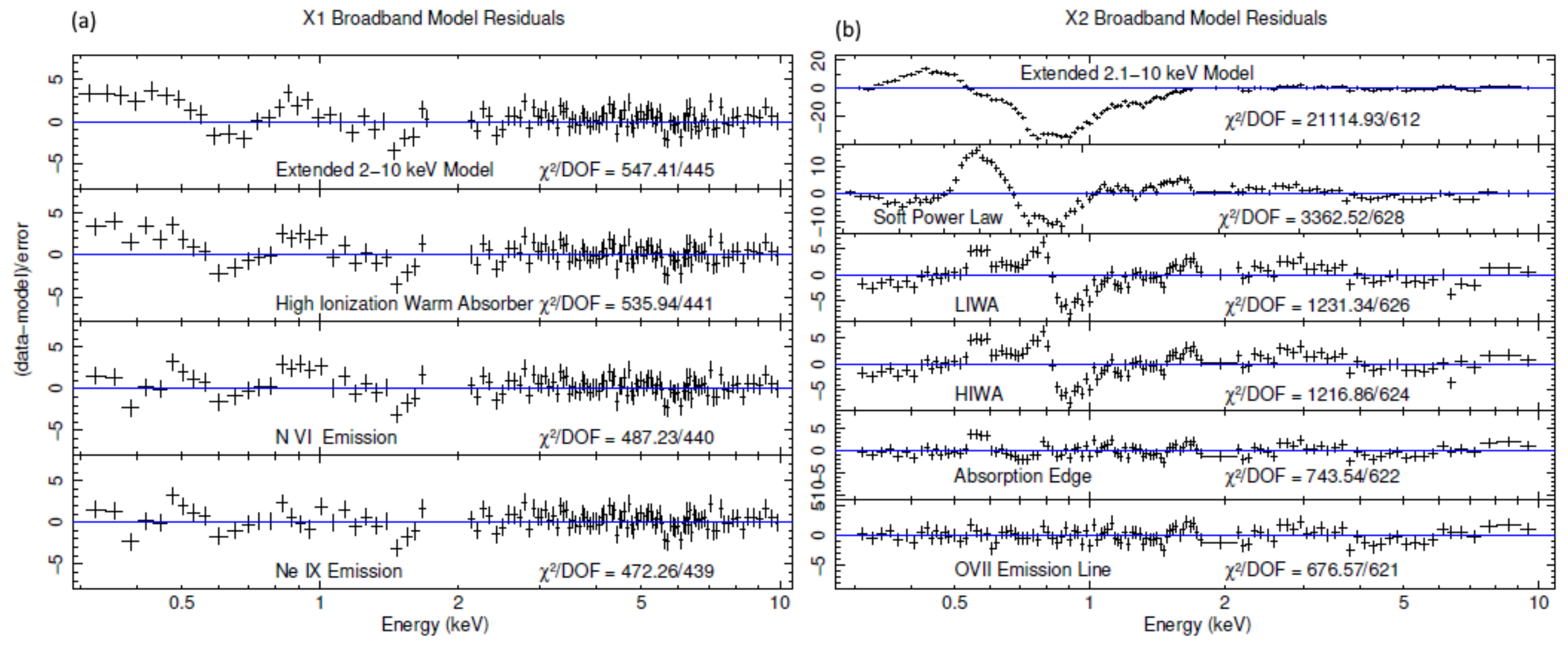}
\caption{X1 (a) and X2 (b) Spectral Fits shown as the ratio of the data to the model.  In X1 after extending the 2-10 keV model to 0.3-10 keV and adding the \replaced{low}{high} ionization warm absorber, the N VI \replaced{emission line is}{and Ne IX emission lines are} included for the best fit. In X2 after extending the \replaced{2}{2.1}-10 keV model to 0.3-10 keV we added the soft power law, the low ionization (LIWA) and high ionization (HIWA) warm absorbers, the absorption edge and the O VII emission line, which achieved the best fit.}
\label{fig:x1x2res}
\end{figure}

Modeled spectra are shown in Figure~\ref{fig:x1x2res} 
for X1 (left panel) and for X2 (right panel).  Model parameters are listed in Table~\ref{tab:xpar}.

\begin{table}[htb!]
\resizebox{0.8\textwidth}{!}{\begin{minipage}{\textwidth}
\caption {Suzaku 0.5-50 keV Model Parameters}
\begin{tabular}{l l c c c c c c} 
\hline \hline
Component & Parameter & 					          S1 & 				          S2 & 				            S3 & 				               S4 & 			                          S5 &				 S6\\
\hline 

Ionized Partial Covering & $N_{H}$ \( \big( 10^{22} cm^{-2}\big) \) & 
										$1.70_{-0.07}^{+0.08}$  & 	$12.86_{-0.36}^{+0.36}$  & 	$6.12_{-0.10}^{+0.11}$  & 	$15.66_{-0.63}^{+0.64}$  & 	$8.36_{-0.09}^{+0.15}$  & 	$11.94_{-0.32}^{+0.32}$ \\
 & Covering Fraction & 							$0.46_{-0.02}^{+0.02}$  & 	$0.81_{-0.01}^{+0.01}$  & 	$0.89_{-0.01}^{+0.01}$  & 	$0.74_{-0.01}^{+0.01}$  & 	$0.88_{-0.01}^{+0.01}$  & 	$0.90_{-0.01}^{+0.01}$ \\
& Log $\xi$ &								$0.30_{-0.20}^{+0.13}$ &	$0.78_{-0.11}^{+0.11}$	&	$0.60_{-0.05}^{+0.05}$	&	$0.30_{-0.10}^{+0.20}$	&	$1.16_{-0.02}^{+0.02}$	&	$1.07_{-0.07}^{+0.03}$				\\
Xillver \footnote{Xillver and Power law normalization are $photons$ $keV^{-1}$$cm^{-2}s^{-1}$ at 1 keV.} & \( \Gamma_{Hard} \) & 					
										$1.50_{-0.01}^{+0.01}$  & 	$1.47_{-0.01}^{+0.01}$  & 	$1.69_{-0.01}^{+0.01}$  & 	$1.43_{-0.01}^{+0.01}$  & 	$1.59_{-0.01}^{+0.01}$  & 	$1.52_{-0.01}^{+0.01}$ \\
 & Log $\xi$ & 								$0.70_{-0.18}^{+0.04}$  & 	$1.83_{-0.06}^{+0.09}$  & 	$1.36_{-0.20}^{+0.17}$  & 	$1.65_{-0.12}^{+0.09}$  & 	$1.65_{-0.09}^{+0.06}$  & 	$1.54_{-0.11}^{+0.14}$ \\
 & Reflection Fraction & 							$0.73_{-0.05}^{+0.05}$  & 	$1.53_{-0.08}^{+0.11}$  & 	$1.23_{-0.07}^{+0.08}$  & 	$2.57_{-0.18}^{+0.19}$  & 	$1.22_{-0.08}^{+0.07}$  & 	$0.96_{-0.08}^{+0.12}$ \\
 & XIS 0 \& 3 Norm. \( \big( 10^{-4} \big) \) & 				$2.06_{-0.01}^{+0.01}$  & 	$1.33_{-0.01}^{+0.01}$  & 	$1.62_{-0.01}^{+0.01}$  & 	$0.69_{-0.01}^{+0.01}$  & 	$1.33_{-0.01}^{+0.01}$  & 	$1.22_{-0.01}^{+0.01}$ \\
& XIS 1 Norm. \( \big( 10^{-4} \big) \) &  				$1.90_{-0.02}^{+0.02}$  & 	$1.26_{-0.01}^{+0.01}$  & 	$1.37_{-0.01}^{+0.01}$  & 	$0.69_{-0.01}^{+0.01}$  & 	$1.29_{-0.01}^{+0.01}$  & 	$1.17_{-0.01}^{+0.01}$ \\
O VII Emission Line \footnote{Gaussian normalization is total $photons$ $cm^{-2}s^{-1}$ in the line of sight.} & XIS 1 Norm. \( \big( 10^{-4} \big) \)& 	
										& & 										$8.37_{-3.47}^{+3.47}$& 	$1.46_{-0.97}^{+0.97}$	 & 	$3.33_{-2.18}^{+2.18}$  & 	$4.24_{-2.36}^{+2.36}$ \\	
Ne IX Emission Line & XIS 0 \& 3 Norm. \( \big( 10^{-5} \big) \)&	&					&					$30.76_{-17.22}^{+14.78}$&	$<4.33$			 & 	$18.41_{-6.86}^{+6.86}$ &					 \\
& XIS 1 Norm. \( \big( 10^{-5} \big) \)&					&					&					$31.68_{-14.78}^{+14.78}$ &	$5.64_{-2.72}^{+2.72}$ & 	$14.61_{-7.90}^{+7.90}$ &					 \\
Ne X  Emission Line & XIS 0 \& 3 Norm. \( \big( 10^{-5} \big) \)&	&					&									&	$2.65_{-1.78}^{+1.78}$ & 		 & \\
& XIS 1 Norm. \( \big( 10^{-5} \big) \)&					&					&					 &					$2.42_{-1.45}^{+1.45}$ & & \\
Fe Absorption Edge & Depth & 						$0.15_{-0.02}^{+0.02}$ & & & & \\
High Ion. Warm Abs. & $N_{H}$ \( \big( 10^{21} cm^{-2}\big) \) & $49.08_{-19.52}^{+30.04}$ &	 $11.09_{-1.72}^{+2.16}$  & 	$31.68_{-12.96}^{+10.34}$  & 				 &		$32.98_{-2.87}^{+3.09}$  & 	$28.98_{-15.70}^{+21.02}$ \\
& Log $\xi $ & 								$2.58_{-0.08}^{+0.08}$ & 	$2.90_{-0.09}^{+0.09}$  &	 $2.93_{-0.08}^{+0.09}$  & 				 &		$2.31_{-0.02}^{+0.02}$  & 	$2.56_{-0.05}^{+0.32}$ \\
Low Ion. Warm Abs. & $N_{H}$ \( \big( 10^{21} cm^{-2}\big) \) & $6.16_{-0.42}^{+0.44}$  & 	$1.65_{-0.17}^{+0.19}$  & 	$3.94_{-0.20}^{+0.21}$  &  	$1.38_{-0.19}^{+0.22}$ & 	$1.66_{-0.17}^{+0.20}$  & 	$2.60_{-0.26}^{+0.29}$ \\
&  Log $\xi$ &								 $1.47_{-0.04}^{+0.04}$  & 	$1.33_{-0.12}^{+0.11}$  & 	$1.06_{-0.08}^{+0.08}$  & 	 $1.25_{-0.41}^{+0.13}$ & 	$0.84_{-0.11}^{+0.11}$  & 	$1.47_{-0.11}^{+0.11}$ \\
Soft Power Law  & \( \Gamma_{Soft} \) & 				$3.31_{-0.11}^{+0.11}$ & & & & &\\
& XIS 0 \& 3 Norm.  \( \big( 10^{-3} \big) \) & 				$1.63_{-0.09}^{+0.09}$ & & & & &\\
& XIS 1 Norm.  \( \big( 10^{-3} \big) \)  & 				$1.73_{-0.09}^{+0.09}$ & & & & &\\
XIS-PIN Calibration & Constant & 					1.154 & 				1.217 & 				1.212 & 				1.206 & 				1.206 & 				1.160 \\
&\(\chi^2 \)/DOF & 								1807.68/1751 & 			1325.03/1317 & 			243.89/219 &			163.07/148  & 			1102.48/1033& 			596.85/562\\
& P-value &									0.1688	&			0.4329 &				0.1928		&		0.1877		&		0.0657		& 		0.1495	\\
\hline 
\end{tabular} 
\end{minipage}}
\label{tab:spar}

\end{table}

\begin{table}[htb!]
\centering
\resizebox{0.8\textwidth}{!}{\begin{minipage}{\textwidth}
\caption{XMM-Newton 0.3-10 keV Model Parameters}
\begin{tabular}
{l l c  c} \hline \hline
Component & Parameter & 						X1 & 					X2  \\
\hline 

Ionized Partial Covering  & $N_{H}$ $\big(10^{22}cm^{-2}\big)$  & $6.30_{-0.17}^{+0.17}$  & 										\tabularnewline
 & Covering Fraction  & 							$0.94_{-0.01}^{+0.01}$  &										 \tabularnewline
& Log $\xi$ &								$0.32_{-0.07}^{+0.07}$ & \\
Full Covering Absorption & $N_{H}$ $\big(10^{20}cm^{-2}\big)$  & 	&					$10.34_{-0.20}^{+0.20}$ \\
Xillver \footnote{Xillver and Power Law normalization are $photons$ $keV^{-1}$$cm^{-2}s^{-1}$ at 1 keV.}  & $\Gamma_{Hard}$   & 
										$1.52_{-0.01}^{+0.01}$  & 		$1.56_{-0.01}^{+0.01}$				\tabularnewline
 & Log $\xi$  & 								$1.30_{-0.16}^{+0.06}$  & 		$0.70_{-0.14}^{+0.14}$	\tabularnewline
 & Reflection Fraction & 							$1.71_{-0.16}^{+0.19}$  & 		$0.69_{-0.04}^{+0.04}$ \tabularnewline
 & pn Norm. $\big(10^{-5}\big)$  & 					$6.44_{-0.07}^{+0.07}$  & 		$26.74_{-0.05}^{+0.05}$		\tabularnewline
Ni K $\alpha$  \footnote{Gaussian normalization is total $photons$ $cm^{-2}s^{-1}$ in the line of sight.}  & pn Norm. $\big(10^{-6}\big)$ &
										$3.94_{-2.31}^{+2.31}$&\\
OVII Emission Line & pn Norm. $\big(10^{-4}\big)$  &  		&						$3.37_{-0.41}^{+0.41}$		 \tabularnewline

Absorption Edge  & Edge Energy $\big(keV\big)$  &  			&				 		$0.84_{-0.01}^{+0.01}$\tabularnewline
 & Depth  &  									&						$0.18_{-0.01}^{+0.01}$\tabularnewline
N VI Emission Line & pn Norm. $\big(10^{-4}\big)$&			$5.62_{-1.30}^{+1.30}$ & \\
Ne IX Emission Line & pn Norm. $\big(10^{-4}\big)$&			$1.05_{-0.35}^{+0.35}$ & \\
High Ion. Warm Abs.  & $N_{H}$ $\big(10^{21}cm^{-2}\big)$ &   	$13.01_{-8068}^{+11.89}$&		$36.51_{-20.76}^{+15.24}$\tabularnewline
 & Log $\xi$  & 								  $3.08_{-0.29}^{+0.49}$&		$3.00_{-0.15}^{+0.58}$\tabularnewline
Low Ion. Warm Abs. & $N_{H}$ $(10^{21}cm^{-2})$ &		  & 						$3.15_{-0.07}^{+0.08}$\tabularnewline
 & Log $\xi$ & 								 &						$1.30_{-0.03}^{+0.03}$\tabularnewline
Soft X-Ray Power Law  & $\Gamma_{Soft}$   & 			& 						$3.83_{-0.02}^{+0.02}$ 		\\
 & pn Norm. $\big(10^{-3}\big)$  & 					 & 						$2.85_{-0.02}^{+0.02}$ 		\tabularnewline
&  \(\chi^2  \)/DOF & 								472.26/438 & 					677.10/623 \\
& P-value &									0.1319 &					0.0656 \\
\hline
\end{tabular}
\end{minipage}}
\label{tab:xpar}
\end{table}

\subsection{\textit{Alternative Models for Soft Excess}}

Extra soft excess emission appears only during the S1 and X2 observations. In Sections 4.1.2 and 4.2.2 it was shown that a steep power law (steeper than the primary power law in the higher energy bands) adequately explains the data. Here additional alternative possible models for this soft excess are considered: blackbody, warm comptonization, thermal bremmstralung, and ionized reflection. 

The redshifted blackbody (BB) model (model ``zbbody")
produced a \replaced{acceptable}{\bf{good}} fit for S1 with $\chi^2$/DOF = \replaced{2071.68/1850}{\bf{1807.13/1755}} and X2 with $\chi^2$/DOF = \replaced{1769.82/1545}{\bf{691.98/605}}.
Our best fits give kT = $\replaced{207.34_{-3.00}^{+3.05}}{\bf{113.57_{-5.37}^{+5.60}}}$ eV for S1 and kT = $\replaced{87.21_{-0.74}^{+0.74}}{\bf{87.78_{-0.74}^{+0.74}}}$ eV
for X2. These values are too high as a more reasonable temperature is closer to kT = 10 eV (see Bechtold et al.
1987, Gierli\'{n}ski \& Done 2004).  We do not consider blackbody to be a physically plausible explanation of the soft excess emission.

Next a warm Compontization model was considered (model ``compST'' Sunyaev \& Titarchuk 1980). In this model the soft X-ray emission is due to Comptonization of lower energy (UV or EUV) seed photons emitted in the outer regions of the accretion disk in the warm atmospheres of the inner regions of the disk.  This model shows a \deleted{reasonaly} good fit for S1 with $\chi^2$/DOF = \replaced{2071.68/1850}{\bf{1813.98/1754}} and an acceptable fit with $\chi^2$/DOF = \replaced{1762.28/1542}{\bf{670.14/602}} for X2.  The fit parameters were $\tau = \replaced{22.68_{-1.97}^{+2.19}}{\bf{22.67_{-0.75}^{+0.19}}}$, kT = $\replaced{355.70_{-33.50}^{+36.13}}{\bf{363.50_{-13.96}^{+14.72}}}$ eV for S1 and $\tau = \replaced{17.08_{-0.10}^{+0.10}}{\bf{14.92_{-0.09}^{+0.09}}}$, kT = $\replaced{362.60_{-4.73}^{+3.49}}{\bf{479.77_{-4.28}^{+4.31}}}$ eV for X2.  

Then a redshifted bremsstrahlung
(model``zbremss'', Kellogg et al. 1975, Karzas \& Latter
1961) was applied to the data.\deleted{For S1 although} \replaced{a}{A} good fit was obtained (with $\chi^2$/DOF = \replaced{2036.31/1849}{\bf{1810.62/1755}}) \added{\bf{for S1}} \deleted{the temperature was too high with kT = $\replaced{296.92_{-27.23}^{+25.99}}{352.29_{-22.6}^{+21.91}}$ eV}\added{\bf{but not for X2 $\chi^2$/DOF =1329.25/625}}.  \deleted{An acceptable fit for X2 was not achieved with $\chi^2$/DOF = 5040.43/3788.}  We do not consider this model further.

Finally an ionized relativistic reflection (IRR) model (model ``reflionx" Ross \& Fabian 2005, Ross et al. 1999) blurred in the Laor model shape (model ``kdblur" Laor 1991) was considered. An \replaced{acceptable}{\bf{good}} fit for S1 was obtained with $\chi^2$/DOF = \replaced{2155.56/1881}{\bf{1800.40/1756}}, but it was not acceptable for X2 with $\chi^2$/DOF = \replaced{1976.26/1543}{\bf{1578.04/623}}.

With these results it appears that warm Comptonization is acceptable as an alternative option \added{\bf{despite the P-values for X2 being inferior to the two power law model}}.  Model parameters are displayed in Table~\ref{tab:alts} for S1 and Table~\ref{tab:altx} for X2.

\begin{table}[htb!]
\centering
\resizebox{0.8\textwidth}{!}{\begin{minipage}{\textwidth}
\caption{S1 Alternate Model Parameters}
\begin{tabular}
{l l c c c c} \hline \hline
Component & Parameter & 							BB Model & 					CompST Model &					Bremss Model &						IRR Model \\
\hline 
Ionized Partial Covering & $N_{H}$ $\big(10^{22}cm^{-2}\big)$ &	$0.78_{-0.0.8}^{+0.09}$ &     			$1.10_{-0.06}^{+0.07}$& 		$1.08_{-0.09}^{+0.07}$&					$2.97_{-0.12}^{+0.13}$ \\
& Covering Fraction &								$0.23_{-0.01}^{+0.01}$ & 			$0.37_{-0.03}^{+0.03}$& 		$0.30_{-0.02}^{+0.02}$&					$0.38_{-0.01}^{+0.01}$\\
& Log $\xi$ &									$0.24_{-0.35}^{+0.22}$ &			$0.29_{-0.26}^{+0.16}$ &		$0.30_{-0.23}^{+0.20}$&									\\
Xillver \footnote{Xillver and Reflionx normalization are $photons$ $keV^{-1}$$cm^{-2}s^{-1}$ at 1 keV.} & $\Gamma_{Hard}$ &	
								 			$1.47_{-0.01}^{+0.01}$& 			$1.51_{-0.01}^{+0.01}$&		$1.48_{-0.01}^{+0.01}$&					$1.59_{-0.01}^{+0.01}$\\
& Log $\xi$ & 									$<1.15$ & 						$0.71_{-0.13}^{+0.28}$& 		$1.02_{-0.30}^{+0.31}$&				 	$0.56_{-0.18}^{+0.13}$\\
& Reflection Fraction & 								$0.59_{-0.04}^{+0.04}$ & 			$0.66_{-0.05}^{+0.05}$&		$0.62_{-0.05}^{+0.05}$&					$0.73_{-0.05}^{+0.07}$ \\
& XIS 0 \& 3 Norm.  $\big(10^{-4}\big)$ &					$2.17_{-0.01}^{+0.01}$ & 			$2.09_{-0.01}^{+0.01}$&		$2.15_{-0.01}^{+0.01}$&	 				$2.18_{-0.01}^{+0.01}$\\
& XIS 1 Norm.  $\big(10^{-4}\big)$ &						$2.03_{-0.01}^{+0.01}$ & 			$1.94_{-0.01}^{+0.01}$& 		$2.00_{-0.01}^{+0.01}$&	 				$2.08_{-0.02}^{+0.02}$\\
High Ion. Warm Abs. & $N_{H}$ $\big(10^{21}cm^{-2}\big)$ &		$81.11_{-10.65}^{+12.06}$ & 			$80.06_{-9.69}^{+10.84}$& 		$74.90_{-12.05}^{+14.16}$&		 			$2.12_{-0.13}^{+0.13}$\\
& Log $\xi$ &									$2.40_{-0.01}^{+0.01}$ & 			$2.38_{-0.01}^{+0.01}$& 		$2.42_{0.01}^{+0.01}$&						$2.33_{-0.02}^{+0.02}$\\
Low Ion. Warm Abs. & $N_{H}$ $\big(10^{21}cm^{-2}\big)$ &		$5.95_{-0.19}^{+0.20}$ & 			$6.13_{-0.44}^{+0.46}$& 		$6.34_{-0.35}^{+0.37}$&	 				$3.96_{-0.13}^{+0.13}$\\
& Log $\xi$ &									$1.44_{-0.02}^{+0.02}$ & 			$1.44_{-0.04}^{+0.05}$& 		$1.41_{-0.04}^{+0.05}$&	 				$1.41_{-0.02}^{+0.02}$\\
Blackbody \footnote{Blackbody normalization is $L_{39}/[D_{10}(1+z)^2]$, $L_{39}$ is luminosity in $10^{39} ergs \ s^{-1}$, $D_{10}$ is distance in 10 kpc} & kT (eV) &	
											$113.57_{-5.37}^{+5.60}$ & &&  \\
& XIS 0 \& 3 Norm. $\big(10^{-5}\big)$ &					$6.22_{-0.71}^{+0.71}$ & &&  \\
& XIS 1 Norm.  $\big(10^{-5}\big)$ &						$6.94_{-0.53}^{+0.53}$ & &&  \\
Comptonized Component \footnote{CompST normalization is $Nf/4$$\pi$$D^2$ where $N$ is the total number of photons, $D$ is the distance, $f = z(z+3)$$y^2$/$\Gamma$$(2x+4)$/$\Gamma$$(z)$, $z$ is the spectral index, $y$ is the injected photon energy in units of temperature, and $\Gamma$ is the incomplete gamma function} & kT (eV) &									&																	$363.50_{-13.96}^{+14.72}$ && \\
& Depth &										&							$22.67_{-0.75}^{+0.19}$ & & \\
& XIS 0 \& 3 Norm.  $\big(10^{-3}\big)$ &					& 							$1.84_{-0.10}^{+0.10}$&  &\\
& XIS 1 Norm.  $\big(10^{-3}\big)$ &						& 							$1.91_{-0.09}^{+0.09}$& & \\
Bremsstrahlung  \footnote{Bremsstrahlung normalization is $3.02 \times 10^{-15} / 4$$\pi$$D^2$ $\int n_en_I$$dV$ where $D$ is the distance in $cm$, and $n_e$ and $n_I$ are the electron and ion densities in $cm^{-3}$} & kT (eV) &
										&							&							$343.46_{-20.04}^{+21.83}$ & \\
& XIS 0 \& 3 Norm. $\big(10^{-2}\big)$ &					 & & 													$1.26_{-0.08}^{+0.08}$& \\
& XIS 1 Norm.  $\big(10^{-2}\big)$ &						& & 													$1.34_{-0.07}^{+0.07}$& \\
Reflionx & $\xi$ &									& &	&																					$48.63_{-5.13}^{+2.73}$ \\
& XIS 0 \& 3 Norm. $\big(10^{-6}\big)$ &					 & & 	&																					$1.47_{-0.53}^{+0.53}$ \\
& XIS 1 Norm.  $\big(10^{-6}\big)$ &						& & 	&																					$1.98_{-0.90}^{+0.90}$\\

& $\chi^2$/DOF &									1807.13/1755 &					1813.98/1754 &				1810.62/1755 &						1800.40/1756			\\
& P-value &										0.1888	&					0.1556 &					0.1735 &							0.1733 \\
\hline
\end{tabular}
\end{minipage}}
\label{tab:alts}
\end{table}

\begin{table}[htb!]
\begin{center}
\resizebox{0.9\textwidth}{!}{\begin{minipage}{\textwidth}
\caption{X2 Alternate Model Parameters}
\begin{tabular}
{l l c  c c c} \hline \hline
Component & Parameter & 							BB Model & 					CompST Model  &					\\
\hline 
Full Covering Absorption & $N_{H}$ $\big(10^{20}cm^{-2}\big)$  & 	$7.28_{-0.22}^{+0.22}$ &			$10.02_{-0.25}^{+0.25}$&			\\
Xillver \footnote{Xillver normalization is $photons$ $keV^{-1}$$cm^{-2}s^{-1}$ at 1 keV.} & $\Gamma_{Hard}$ &
											$1.61_{-0.01}^{+0.01}$ &			$1.60_{-0.01}^{+0.01}$&			\\
& Log $\xi$ &									$0.30_{-0.10}^{+0.04}$ &			$0.81_{-0.29}^{+0.22}$ &										\\
& Reflection Fraction &								$0.75_{-0.04}^{+0.04}$ &			$0.67_{-0.04}^{+0.04}$&											\\
 & pn Norm. $\big(10^{-4}\big)$  & 						$1.78_{-0.01}^{+0.01}$  & 			 $2.55_{-0.01}^{+0.01}$&											\\
OVII Emission Line \footnote{Gaussian normalization is total $photons$ $cm^{-2}s^{-1}$ in the line of sight.} & pn Norm. $\big(10^{-4}\big)$  &  	
											$1.29_{-0.30}^{+0.30}$&			$2.93_{-0.41}^{+0.41}$&										\\
Absorption Edge & Energy (keV) &						&							$0.84_{-0.01}^{+0.01}$&										\\
& Depth &										&							$0.17_{-0.01}^{+0.01}$&								\\
High Ion. Warm Abs. & $N_{H}$ $\big(10^{21}cm^{-2}\big)$ & 		$ 2.35_{-0.18}^{+0.16} $ &			$ 8.17_{-6.58}^{+8.63}$&												\\
& Log $\xi$ &									$ 2.05_{-0.02}^{+0.04} $ &			$ 2.54_{-0.13}^{+0.72}$&												\\
Low Ion. Warm Abs. &  $N_{H}$ $\big(10^{21}cm^{-2}\big)$ &		$2.20_{-0.07}^{+0.05}$	&			$3.23_{-0.07}^{+0.07}$&									\\
& Log $\xi$ &									$1.28_{-0.02}^{+0.02}$ &			$1.27_{-0.03}^{+0.03}$&								\\
Blackbody  \footnote{Blackbody normalization is $L_{39}/[D_{10}(1+z)^2]$, $L_{39}$ is luminosity in $10^{39} ergs \ s^{-1}$, $D_{10}$ is distance in 10 kpc}  & kT (eV) &		
											$87.78_{-0.74}^{+0.74}$ &			&							&						\\
& pn Norm    $\big(10^{-4}\big)$ &						$1.85_{-0.02}^{+0.02}$ &			&							&							\\
Comptonized Component \footnote{CompST normalization is $Nf/4$$\pi$$D^2$ where $N$ is the total number of photons, $D$ is the distance, $f = z(z+3)$$y^2$/$\Gamma$$(2x+4)$/$\Gamma$$(z)$, $z$ is the spectral index, $y$ is the injected photon energy in units of temperature, and $\Gamma$ is the incomplete gamma function} & kT $(eV)$ &								&																		$479.77_{-4.28}^{+4.31}$&			&							\\
& Optical Depth &									&							$14.92_{-0.09}^{+0.09}$&			&								\\
 & pn Norm. $\big(10^{-3}\big)$  & 						 &				 			 $2.37_{-0.02}^{+0.02}$&			&								\\
& $\chi^2$/DOF &									691.98/605 &					670.14/602 &													\\
& P-value &										0.0080 &						0.0279 \\
\hline
\end{tabular}
\end{minipage}}
\label{tab:altx}
\end{center}
\end{table}

\subsection{\textit{Luminosity and Model Parameter Relation}}

Once the broadband models for all of the observations were formed, analysis was carried out on how model parameters changed with the \added{\bf{intrinsic}} luminosity of the source. Intrinsic luminosity was estimated using the unabsorbed 0.5-10
keV luminosity reported by the XSPEC ``lumin'' command. \added{\bf{Henceforth all references to luminosity refer to the intrinsic luminosity}}.  The emission line intensity did not show any clear correlation with luminosity.

\begin{figure}[htb!]
\includegraphics[scale=1]{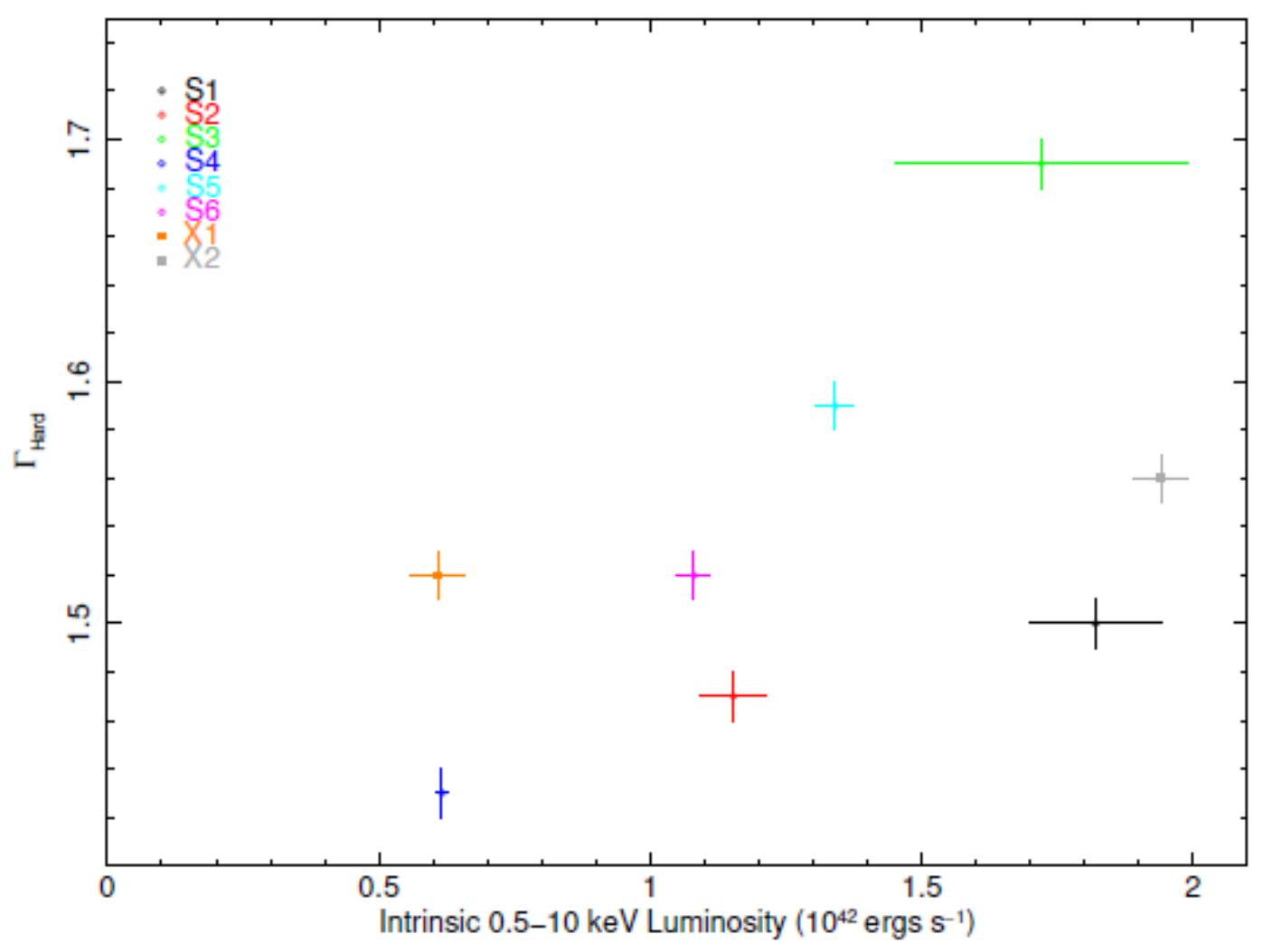}
\caption{Hard Photon Index vs \added{\bf{Intrinsic}} Luminosity. We note a \added{\bf{weak}} positive correlation.}
\label{fig:pogam}
\end{figure}

\begin{figure}[htb!]
\includegraphics[scale=1]{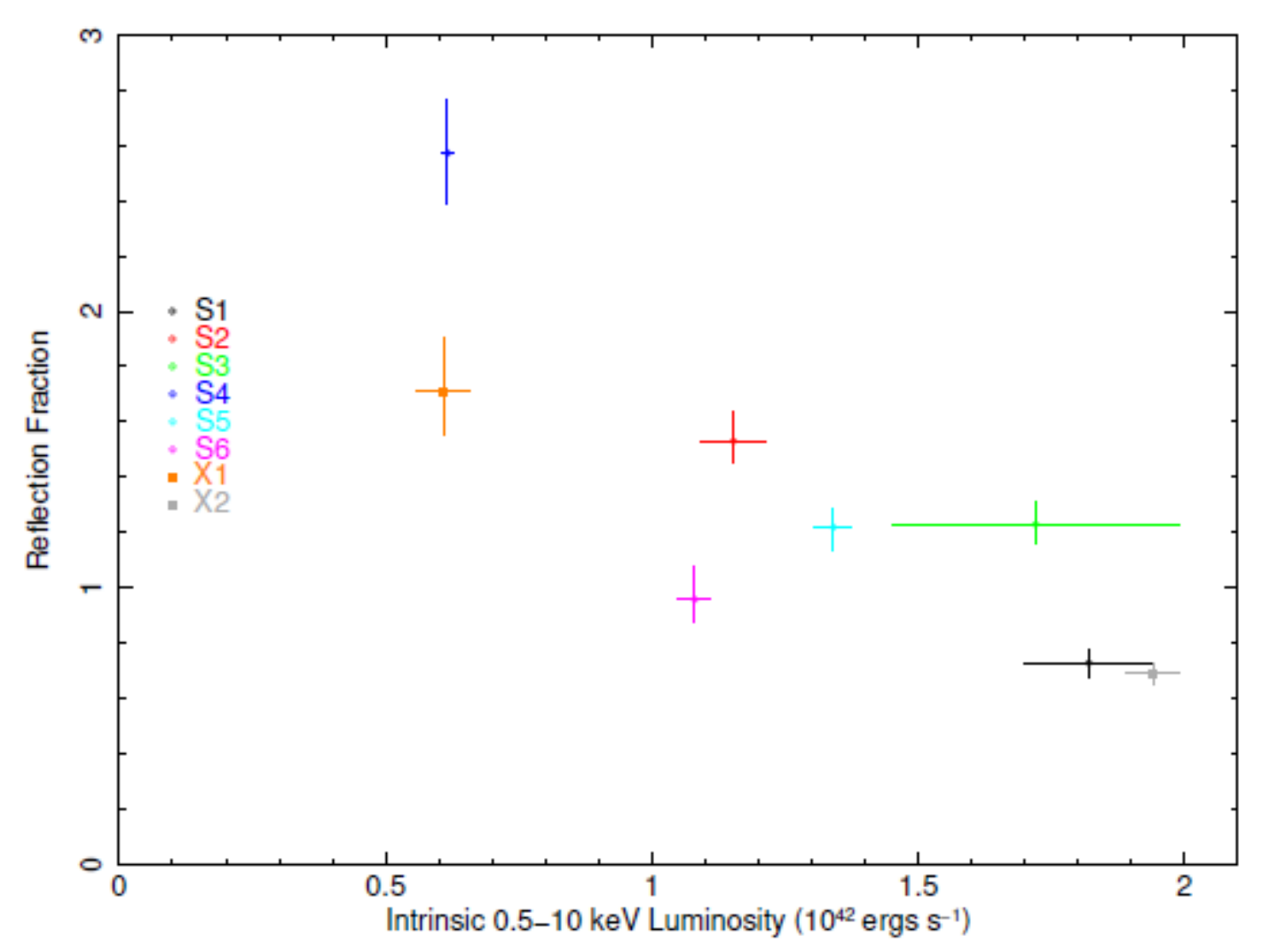}
\caption{Reflection Fraction vs \added{\bf{Intrinsic}} Luminosity. There is a negative correlation.}
\label{fig:lumref}
\end{figure}

\begin{figure}[htb!]
\includegraphics[scale=1]{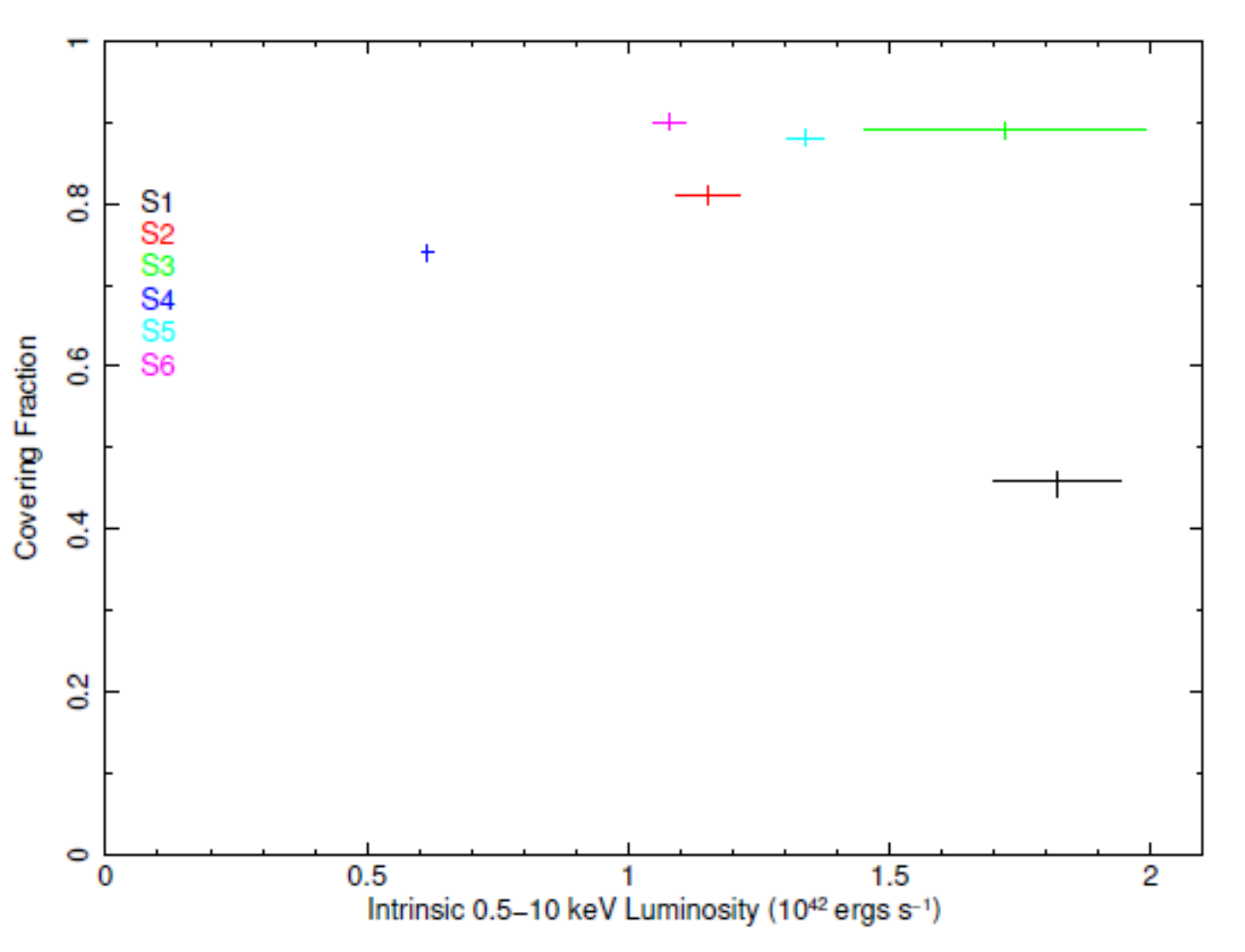}
\caption{\deleted{Neutral}Partial Covering Fraction vs \added{\bf{Intrinsic}} Luminosity. The covering fraction is nearly constant for less bright states, while it drops significantly for the brightest S1 state.}
\label{fig:pccf}
\end{figure}

\begin{figure}[htb!]
\includegraphics[scale=1]{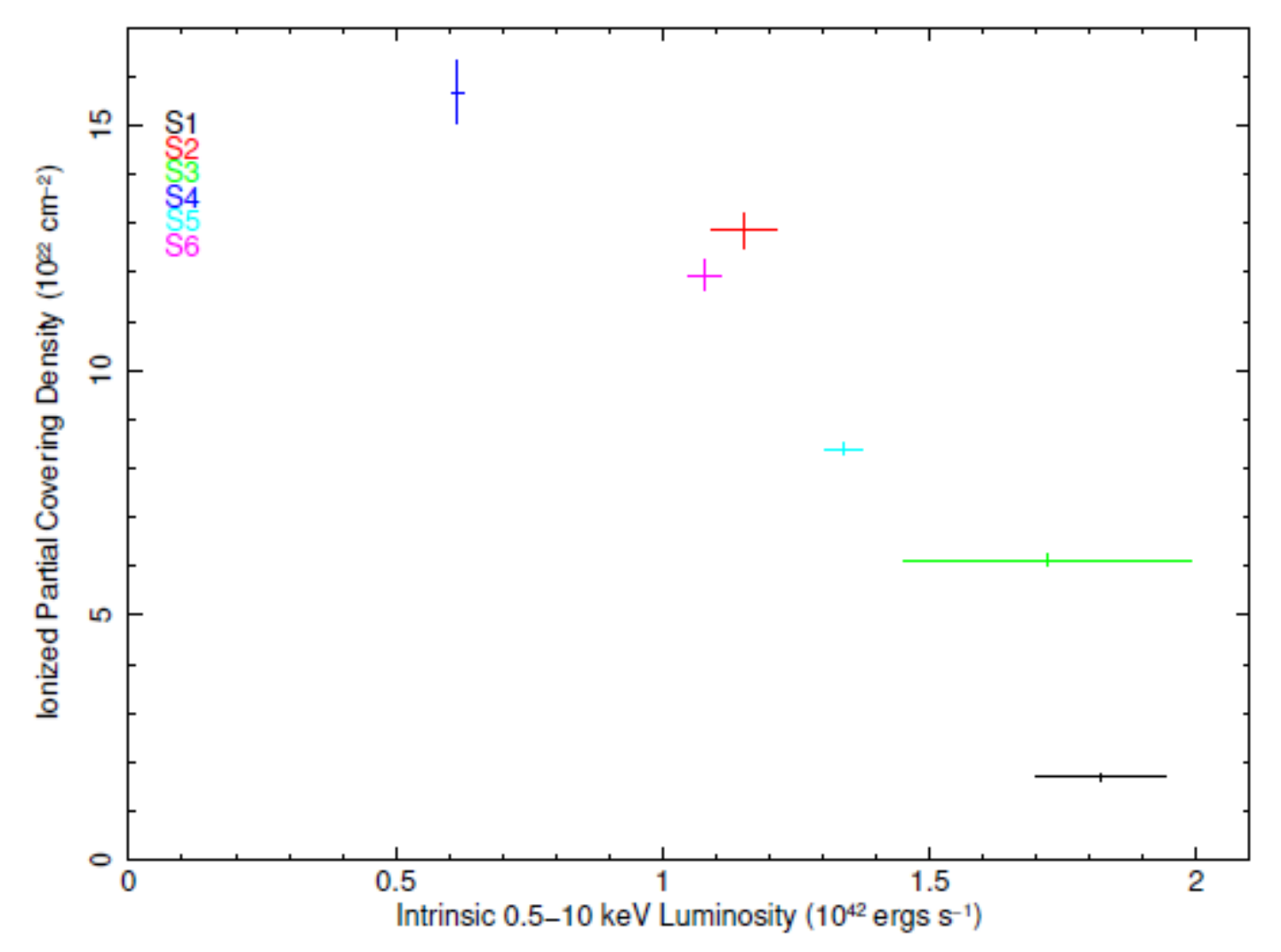}
\caption{\deleted{Neutral}Partial Covering Column Density vs \added{\bf{Intrinsic}} Luminosity. The column density generally decreases with luminosity.}
\label{fig:pcnh}
\end{figure}

\begin{figure}[htb!]
\includegraphics[scale=1]{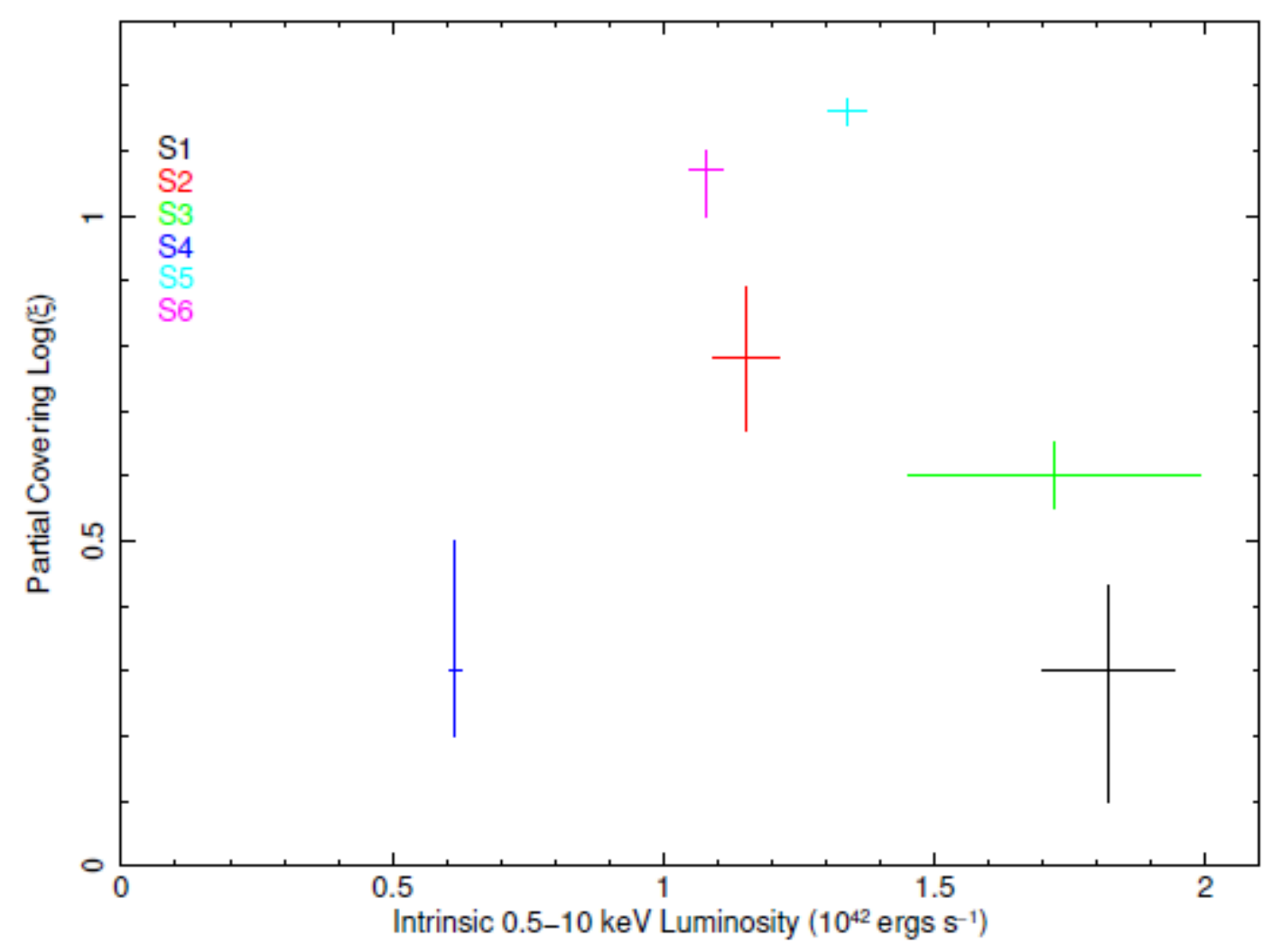}
\caption{Partial Covering Ionization Parameter vs \added{\bf{Intrinsic}} Luminosity. There is no obvious trend.}
\label{fig:pcip}
\end{figure}

Here we would like to present some parameters which pose clear correlations with luminosity. \replaced{First a correlation with
luminosity is apparent (see Figure~\ref{fig:pogam}) for the photon index of the primary 2-10 keV power law emission. 
It slowly increases with luminosity. }{\bf{The hard photon index displayed a weak correlation with luminosity.  On average the higher luminosity states had a higher index than the lower ones although S1 had a markedly low index for its luminosity, comperable to X1 which was one of the lowest luminosity states.}}
\added{\bf{Figure~\ref{fig:lumref} displays the reflection fraction against luminosity.  There is a general negative trend implying reflection is a stronger component for the lower luminosity observations.}}
\added{\bf{Figures~\ref{fig:pccf}-~\ref{fig:pcip} plot components of the ionized partial coverer against luminosity for the \textit{Suzaku} observations.}}  \deleted{In Figure 10 partial covering fraction is plotted against luminosity. The fraction seems to be similar and higher among five \textit{Suzaku} observations S2 - S6 in the lower flux phase, while it is very low for S1 in the bright phase.} In general, Markowitz et al. (2009) in their warm absorber analysis report that from their velocity data the high and low ionization warm absorber clouds are located at around the broad line region (BLR) and narrow line region (NLR). Cold absorbers are expected to be further away. Therefore it is unlikely that the partially covering clouds have changed significantly over the six \textit{Suzaku} observation periods which were taken only within several weeks span.  \textit{XMM-Newton} observations on the other hand were taken a few years apart from each other and the \textit{Suzaku} 
observations, which is enough time for absorption features to change.  \added{\bf{For this reason the \textit{XMM-Newton} observations are absent from Figures~\ref{fig:pccf}-~\ref{fig:pcip}.}}
\deleted{Although the \textit{XMM-Newton} observations did not need to be consistent with the \textit{Suzaku} ones, the covering fraction for X1 in the \textit{XMM-Newton} dim state was similar to the \textit{Suzaku} results in the lower luminosity states S2 to S6.}  
\added{\bf{In Figure \ref{fig:pccf} partial covering fraction is plotted against luminosity. The fraction seems to be similar and higher among five \textit{Suzaku} observations S2 - S6 in the lower flux phase, while it is very low for S1 in the bright phase.  }}The column density of the \textit{Suzaku} observations showed a negative correlation
with the luminosity which is shown in Figure~\ref{fig:pcnh}. 
\deleted{The \textit{XMM-Newton} observations also showed this negative correlation.} \added{\bf{The same treatment was applied to the ionization parameter of the partial coverer (Figure~\ref{fig:pcip}) but no trend was visible.}}  The \textit{Suzaku} trend will be discussed further in Section 5.1.   

\begin{figure}[htb!]
\includegraphics[scale=1]{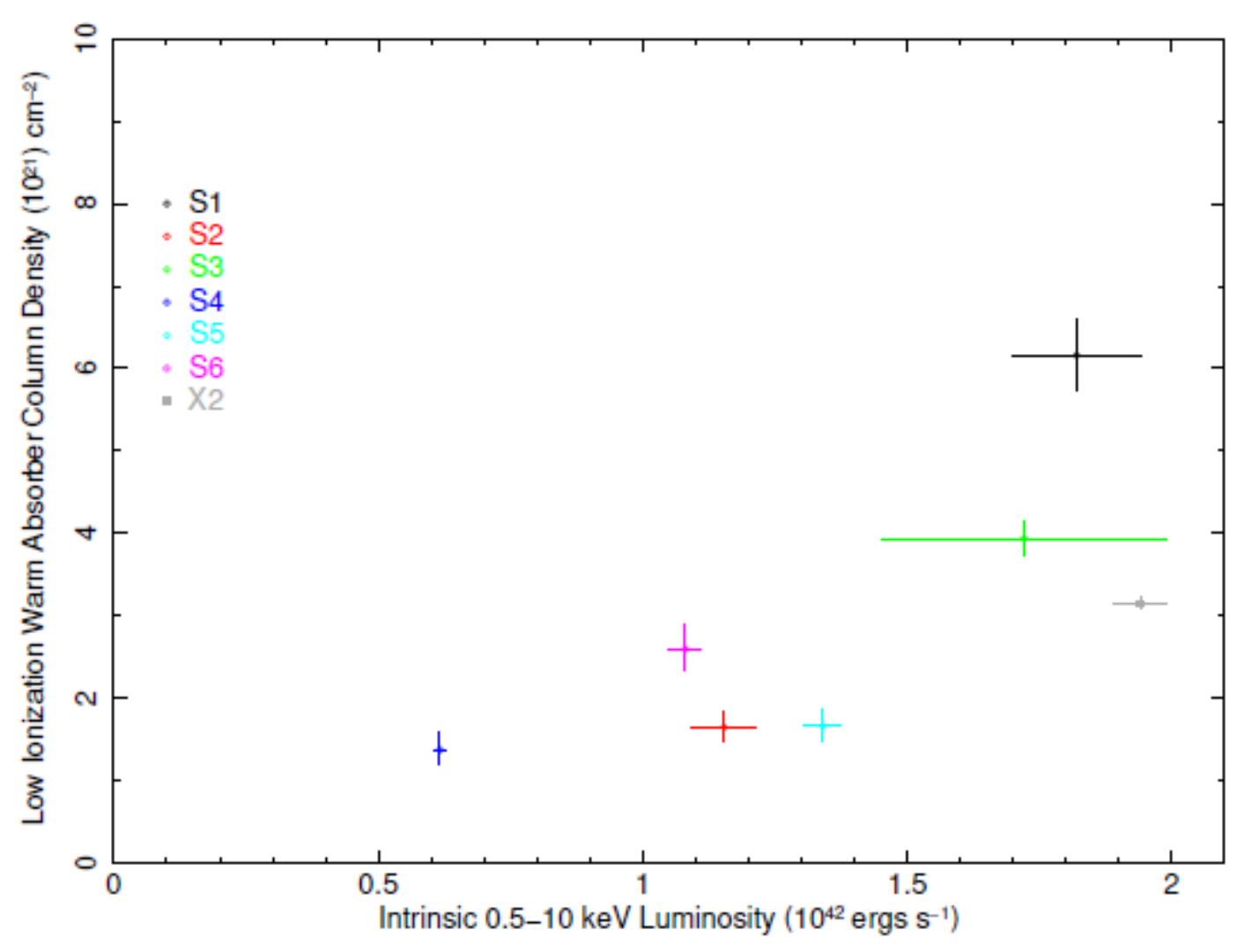}
\caption{Low Ionization Warm Absorber Column Density. We note a positive trend.}
\label{fig:linh}
\end{figure}

\begin{figure}[htb!]
\includegraphics[scale=1]{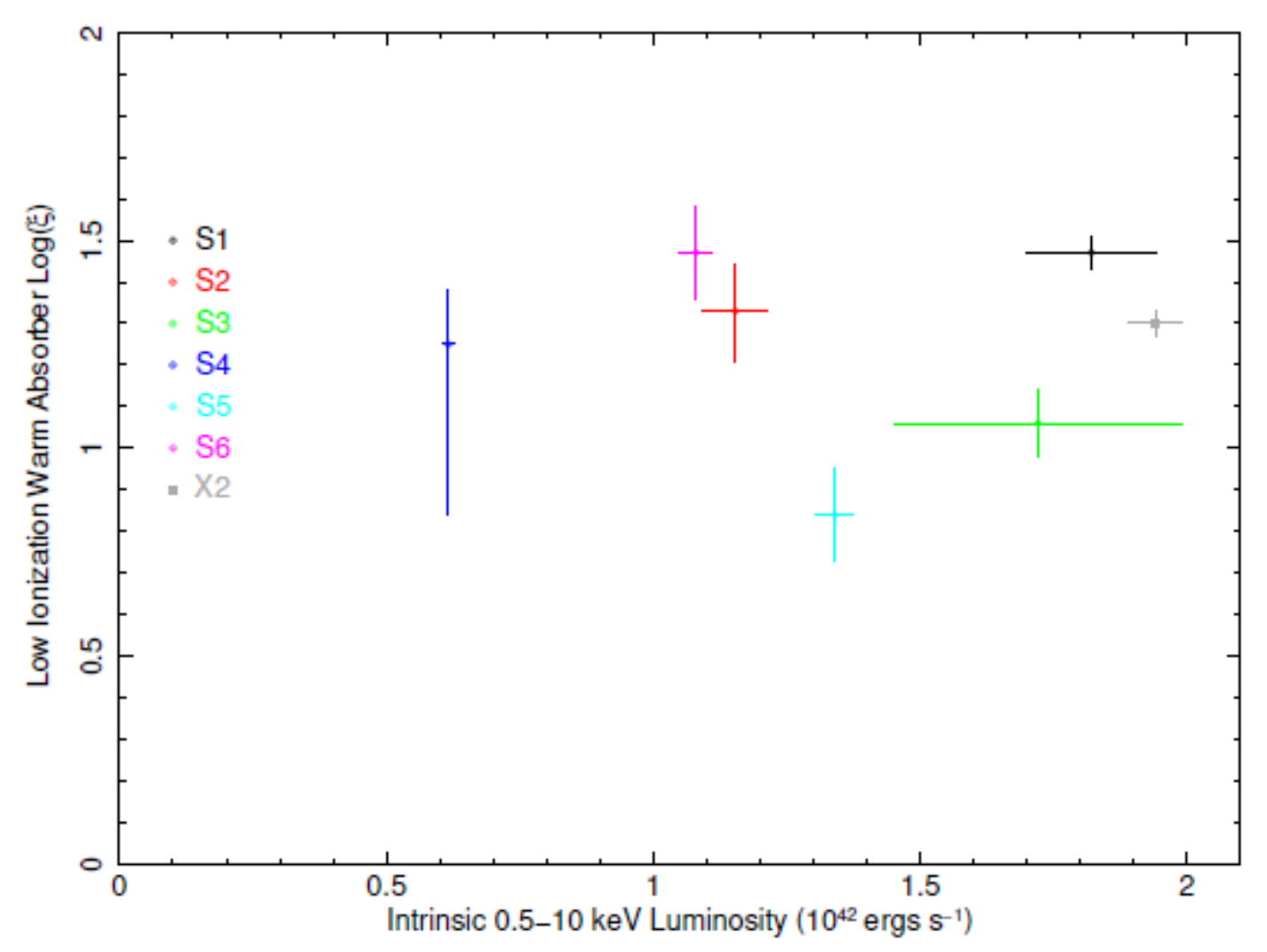}
\caption{Low Ionization Warm Absorber Ionization Parameter.  No clear trend is present.}
\label{fig:liip}
\end{figure}

\begin{figure}[htb!]
\includegraphics[scale=1]{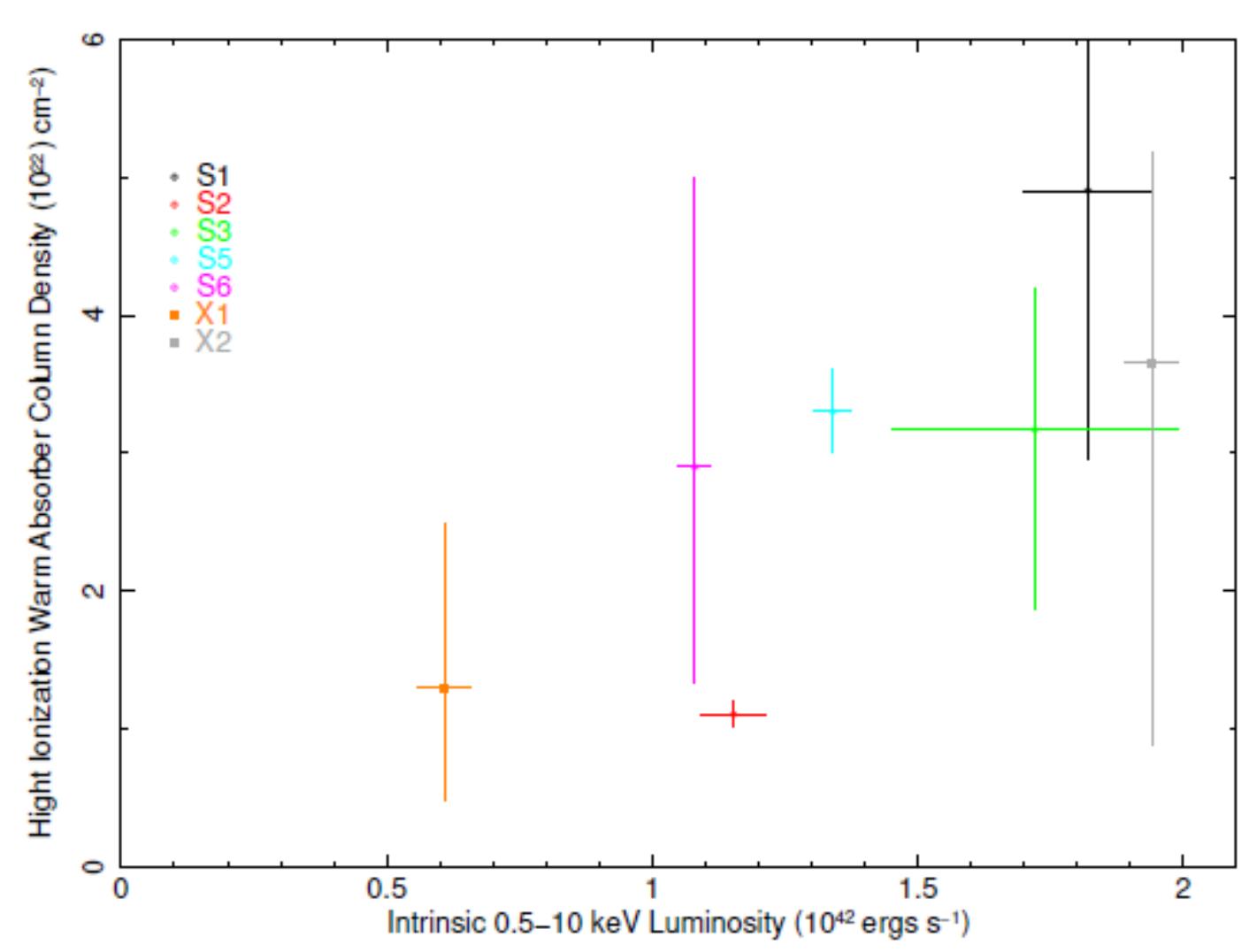}
\caption{High Ionization Warm Absorber Column Density.   No clear trend is present.}
\label{fig:hinh}
\end{figure}

\begin{figure}[htb!]
\includegraphics[scale=1]{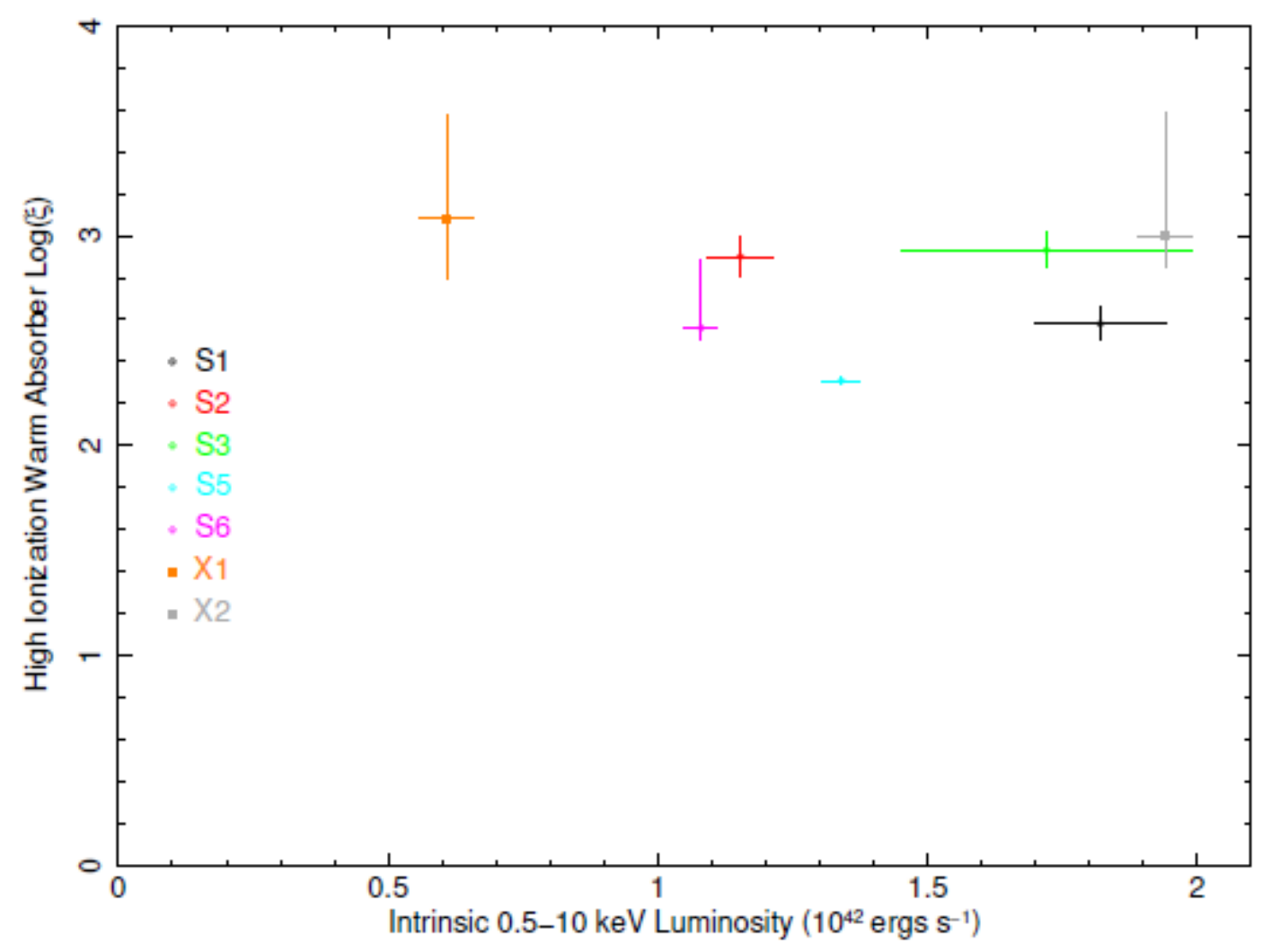}
\caption{High Ionization Warm Absorber Ionization Parameter.  No clear trend is present.}
\label{fig:hiip}
\end{figure}

\added{\bf{Figures~\ref{fig:linh}-~\ref{fig:hiip} display the parameters of the warm absorbers (column density and ionization parameter) against luminosity.  The column density of the low ionization warm absorber (Figure~\ref{fig:linh}) showed a positive correlation with luminosity.  However, the ionization parameter of that absorber (Figure~\ref{fig:liip}) as well as both the column density (Figure~\ref{fig:hinh}) and ionization parameter (Figure~\ref{fig:hiip}) of the high ionization warm absorber showed no obvious trend.  }}

\section{DISCUSSION}

\subsection{\textit{The Unified Model}}

Referring to the flux-flux plots in section 3.2 (Figures~\ref{fig:scc} and~\ref{fig:xcc}) the lower flux states show a
steep slope for our chosen bands while high flux states show a more gradual slope. 
Note that only S1 and X2 are in the bright branch where soft excess is dominant. As the hard band (\replaced{2}{2.1}-10 keV) increases to about twice as high, the soft band (0.5 - \replaced{2}{1.7} keV) goes up about five times as high (see Figure~\ref{fig:scc}). 
The best fit spectral models achieved in Sections 4.1.2 and 4.2.2 indicate that only S1 and X2 have soft excess, which is consistent with our flux-flux analysis.

Based on these results we propose a unified model which is consistent with all eight observations by the two satellite missions.  In this model in the lower states (\textit{Suzaku} S2 to S6 and \textit{XMM-Newton} X1) the intrinsic radiation directly from the central power house is the hard power law emission from a hot corona above an accretion disk, which is the conventional Compton model for Seyfert nuclei.
This interpretation is consistent with what Yang et al. (2015) found in their studies of correlation between the photon index and X-ray luminosity. Their results show that the photon index increases with luminosity when the X-ray emission comes from the corona in the disk-corona model, and moreover that for fainter objects within this class the power law index can be as low as $\sim$ 1.4 - 1.7 (see their Figure 2). The source NGC 3227 is a relatively low luminosity ($\sim 10^{42} ergs \ s^{-1}$) Seyfert where the photon index of the hard X-ray power law emission is rather low, around \replaced{1.27}{1.4} - \replaced{1.6}{1.7}\deleted{, and it increases with luminosity (see our Figure~\ref{fig:pogam})}.

This hard power law emission continues to the high states (S1 and X2). However, in these high states the total emission is dominated by an additional strong highly variable soft excess.
We identify the location of this soft excess emission tentatively as the warm atmospheres of the accretion disk. The data of this component is consistent with an additional steep power law with $\Gamma$ $\sim$ \replaced{3.5-3.7}{3.3-3.85}. Such power law emission in the atmosphere of the accretion disk can be produced through magnetic effects, such as through microflares like in the solar magnetosphere just above the surface. As the accretion rates and hence luminosity increase it is possible that such additional events are triggered due, e.g., to magnetic instability, on the atmospheres just above the disk surface. Short timescales of flux increase and variability of the soft excess may be due to the timescale of magnetic instability. Another possibility will be warm Comptonization. The analysis carried out in Section 4.3 shows that the warm Compton model is also consistent with the data.  Using the time scale of time variability to estimate the size of the emission region, we calculate the size as $\leq 10^{15} cm$ ($80 R_g$).  In this scenario the soft excess is caused by Comptonization of softer photons from the outer regions of the disk in the warm atmospheres of the accretion disk closer to the center of the disk. 

These primary continuum emissions from the central power house are absorbed by various cold and warm materials further out in the line of sight. There is also an additional reflection component as evidenced by the narrow Fe K$\alpha$ emission.   
Consider the relationship between the behavior of partially covering absorbers and luminosity
found for the \textit{Suzaku} observations in Subsection 4.4 (see Figures~\ref{fig:pccf} and~\ref{fig:pcnh}). 
The column density of the partially covering absorber generally decreased with luminosity (see Figure~\ref{fig:pcnh}).
Furthermore S1 in a brightest state has a significantly lower covering fraction than in the lower states (see Figure~\ref{fig:pccf}). 
The \deleted{cold} partially covering absorbers \added{\bf{and fully covering neutral absorber (X2)}} should be located further away than \replaced{warm}{fully covering} ionized absorbers which were found to be located at least as far away as the BLR while some are in the NLR, from the velocity data noted in Markowitz et al. (2009). As there are years between the individual \textit{XMM-Newton} observations and \textit{Suzaku} observations, the partial covering absorber can be different for X1, X2, and the \textit{Suzaku} observations.  A timescale of years is enough time for the clouds to drift out of line of sight.  However, there is only about a week between individual \textit{Suzaku} observations, thus the same partial covering absorbers are most likely obscuring the central emissions during the six \textit{Suzaku} observations.

Therefore we offer the following explanation.
The size of the emission region increases with luminosity in a similar
way as described by Haba et al. (2008) for NGC 4051. In this scenario the partially covering cloud
is lumpy and moreover denser near the center above the primary emission region. Due to this
the lowest flux observation (S4) when the primary emission region is small requires both a high covering fraction and high column density as found. As the source becomes brighter the size of the emission region increases. The emission region is
still obscured but the outer parts of it are absorbed by a less dense portion
of the outer parts of the clouds. This can still yield a relatively high covering fraction (e.g. S3) but should yield a lower average column density (both S3 and S1) than S4. In the highest luminosity
state (S1) the outermost parts of the emission site has become so extended that they are
hardly obscured by the cloud. The part of the emission region that
is still obscured is mostly covered by the lower density outer regions of the absorber.
This yields a low covering fraction and a low column density for the absorber in S1.
In this model the size of the corona where the primary power law is emitted increases with luminosity. 
That is consistent with other work on the size of the corona by, e.g. Kara et al. 2019.

\subsection{\textit{Comparison With Other Work}}

Gondoin et al. (2003) analyzed the 2000 \replaced{XMM}{\textit{XMM-Newton}} data of NGC 3227. The spectrum above 4 keV is well fitted by a hard power law continuum with $\Gamma$ $\sim$ 1.5 and an absorption edge at 7.6 keV. In addition a narrow Fe K emission line is detected at 6.4 keV. The continuum is heavily absorbed at soft band by dense neutral gas with $N_H = 6.6_{-0.1}^{+0.1} \times 10^{22} cm^{-2}$ covering $\sim$ 90\% of the central source. The soft continuum is also attenuated due to ionized material with $N_H = 8.9_{-0.9}^{+0.9} \times 10^{21} cm^{-2}$. They also noted variability in the continuum emission with a few ks timescale. Our analysis of this observation (X1) agrees with their results although their warm absorber had a higher density. 

Markowitz et al. (2009) presented results of a 100 ks 2006 \replaced{XMM}{\textit{XMM-Newton}} observation of NGC 3227. Their best-fit model to the EPIC pn spectrum consists of a moderately flat hard X-ray power law with $\Gamma$ of 1.57 absorbed by cold gas with $N_H = 2.9_{-0.8}^{+0.3} \times 10^{21}$ $cm^{-2}$, and a strong soft excess with steep power law with $\Gamma$ of 3.35. Both were absorbed by cold gas with $N_{H} = 8.7_{-0.5}^{+0.6} \times 10^{20} cm^{-2}$.  The hard X-ray power law was consistent with the standard disk-corona model, although these authors commented that the power law index was rather low. These authors discussed the possible origin of the soft excess, but did not come up with a definite physical model. They find the data to be consistent with the warm Compton model also, but noted that the variability behavior of the soft excess is different from the UV variability. In this model the UV seed photons coming from regions further away are supposed to be comptonized in the warm atmosphere of the disk closer to the center. They also commented on a possibility of a jet as the origin if this component is steep power law, but then they noted that this source does not have any radio jets.

 We carried out independently the analysis of multiple observations by two different missions in the broad energy range from \replaced{0.5}{0.3} to 50 keV, and found a common model which can explain all eight observations in various flux levels. On the other hand Markowitz et al.(2009), as well as others, studied only one or a limited number of our observation targets. The main focus of our studies is on the central power house itself, while Markowitz et al. (2009) concentrated mostly on the detailed studies of the effects of the surrounding material.
Our unified model is consistent mostly with the model by Markowitz et al. (2009), although there are some minor differences.
For instance, our results from a rather simple warm absorber model gave somewhat thicker high ionization absorber.
As stated in Section 5.1 we find that a rather low index value for the hard power law is acceptable. As to the nature of the soft excess Markowitz et al. (2009) did not specify any definite physical model. Our suggestion is that the steep power law could be due to some magnetic activity in the warm atmosphere above the accretion disk.
If it is warm Comptonization, our suggestion is that the seed photons could be from the extreme ultraviolet (EUV) region closer to the center than ultraviolet (UV). This interpretation can avoid the conflict with the  possible discrepancy between soft excess and UV variability reported by Markowitz et al. (2009).

 The major focus of Markowitz et al. (2009) is the detailed analysis of absorbing material, especially the warm absorbers, by utilizing both EPIC and RGS. They found two absorbing layers with similar column densities of $\sim 1-2 \times 10^{21} cm^{-2}$ but with different ionization states, one at a high state with $log$ $\xi_{hi} = 2.93_{-0.09}^{+0.15}$ $erg$ $cm$ $s^{-1}$ and another at a low state with $log$ $\xi_{lo} = 1.45_{-0.07}^{+0.16}$ $erg$ $cm$ $s^{-1}$. The outflow velocities are detected with $2060_{-170}^{+240}$ $km$ $s^{-1}$ and $420_{-190}^{+430}$ $km$ $s^{-1}$ for the high and low ionized absorbers, respectively. This information gives the estimated location of the high ionization clouds at the BLR and the NLR for the low ionization clouds. 

Noda et al. (2014) studied all six of \textit{Suzaku} 2008 observations of NGC 3227 in the energy range from 2 to 50 keV. In their model the primary continuum emission in the lower luminosity states S2 to 
S6 is absorbed flatter power law emissions with $\Gamma$ $\sim$ 1.6, while the most luminous state S1 is dominated by less absorbed steeper power law continuum with $\Gamma$ $\sim$ 2.3 although a weaker flatter power law still exists. In addition there is a cold reflection component which is evidenced by the narrow Fe K$\alpha$ line which appears all through the six observations.
Their interpretation is that the source undergoes a low to high state transition similar to the case of stellar mass black holes. In this model as the accretion rate and hence luminosity increase the disk system physically changes from an optically thin, geometrically thick ion torus to an optically thick, geometrically thin disk with a corona at a critical accretion rate. The observations S2 to S6 in the lower states belong to the torus system while S1 is in a high state with the disk-corona system where the torus still exists but it is greatly diminished only to the central region. The corona emits the steep power law while the torus is responsible for the flatter power law.   

The Noda et al. (2014) model is substantially different from our unified model and also the model presented by Markowitz et al. (2009).
\replaced{It could be because their analysis is confined to the 2-50 keV ranges.  For instance, their model does not explain the soft excess which is significant below 2 keV for S1. In the high luminosity state S1 their primary hard X-ray continuum is dominated by steep power law,
while in the model by Markowitz et al. (2009) and our current unified model it consists of one flatter power law.
When their model was extended to below 2 keV it did not yield acceptable fits to the soft band which is the major component for both S1 and X2. }{{\bf In the high luminosity state S1 their primary hard X-ray continuum is dominated by steep power law,
but both in the model by Markowitz et al. (2009) and ours in the high state it consists of one flatter power law.
The main reason is that their analysis is confined to the 2-50 keV ranges. However, the effects of warm absorption, various emission features and soft excess are crucial mostly in the lower bands below 2 keV, and therefore the model obtained by excluding these low energy bands will fail to give a better understanding of the source. When their model was extended to below 2 keV it did not yield acceptable fits in these soft bands.

More recently NGC 3227 was observed by \textit{XMM-Newton} and \textit{NuSTAR} for six times during 2016 Nov. 9 - 2016 Dec. 9 and additionally by \textit{NuSTAR} on 2017 Jan. 21. Lobban et al. (2020) carried out X-ray valiability analysis of the data from these observations, while Turner et al. 
(2018) concentrated on the last two observations of the 2016 campaign. During all these observations NGC 3227 was in the bright state. Their primary emission is a power law with $\Gamma$ of 2 in Turner et al. (2018) while 1.4 - 1.7 in Lobban et al. (2020), with a black body (Turner et al. 2018) or Comptonized disk blackbody (Lobban et al. 2020) added as a soft excess. The primary emission model by Lobban et al. (2020) is essentially consistent with our model at the bright state, while the Turner et al. (2018) model is somewhat simpler. Both of these authors agree with our finding on the primary emission from the center: (i) softer when brighter behavior of the power law component, (ii) the bulk of the short variability being continuum driven, and (iii) presence of a strong variable soft excess.

On the other hand, their major focus is the effects of the reprocessing surrounding gas, which is responsible for additional complicated absorption and emission features mostly caused in the BLR but some in the NLR clouds, as well as the neutral reflection from material further away. But some of the reprocessing material are outflow winds from regions closer, e.g., near the accretion disk. Lobban et al. (2020) analyzed all of the \textit{XMM-Newton} and \textit{NuSTAR} data from the 2016 campaign. These clouds are mostly observed as outgoing winds. There are three zones for the warm absorbers with different ionization states. These authors found a strong low frequency hard lag and evidence for a soft lag at higher frequencies. They may be considered to arise from outflowing disk winds via energy-dependent scattering. Turner et al. (2018) mainly investigated the large dip in the light curve exhibited near the end of the 2016 observations, which was also identified by Lobban et al. (2020). It is proposed that this dip is due to a cloud passing through the line of sight in the region near the inner edge of the BLR. In conclusion these papers concentrated mostly on variability in the order of days which are produced by the clouds further away from the central primary source. 

These authors did not discuss the shorter (less than $\sim$ days) variability originating in the central source close to the black hole, while the major focus of our studies, on the other hand, is the central source. However, our current studies of the earlier observations also included the effects of the surrounding absorbing/emitting and reflecting material. The new observations took place much later, $\sim$ 10 years, after the observations treated in the current paper, but our preliminary study shows that our current model is consistent with the newer observations also. In a subsequent paper we will report in detail our spectral and temporal analysis of the data from these new observations.}}

\section{Summary and Concluding Remarks}

We carried out the time-averaged spectral analysis of the combined data for NGC 3227, from two \textit{XMM-Newton} and six \textit{Suzaku} observations. A unified model was constructed which is consistent with all of these observations in a broad energy band from \replaced{0.5}{0.3} to 50 keV. It consists of a hard power law continuum emission covered by cold and warm absorbers and cold reflection. It comes from a corona above an accretion disk where softer photons from the colder disk is Comptonized by hot electrons in the corona. During the bright states an additional soft excess modeled by a steeper power law or warm Comptonization appears. It dominates behavior below 2 keV. This component is highly variable and its behavior is complex. Therefore, the detailed studies of its behavior, such as its short variability in luminosity and spectra, will be carried out separately in a subsequent paper where time-resolved analysis will be applied. 

\acknowledgments{This work was supported by NASA Grant NNX16AF32G. We thank the referee for valuable suggestions which helped to significantly improve this paper. S. Tsuruta thanks IPMU and WPI for their support where this work was partly carried out.  This research is based on observations obtained with the \textit{Suzaku} satellite, a joint mission by Japan (JAXA) and the USA (NASA).  This work also utilizes observations made with the \textit{XMM-Newton} satellite, an ESA X-ray mission with instruments and contributions directly funded by ESA member states and NASA.  This has made use of HEASARC online services, supported by NASA's Goddard Space Flight Center (GSFC).  This research has also used the NASA/IPAC Extragalactic Database (NED), which is operated by the Jet Propulsion Laboratory, California Institute of Technology, under contract with NASA. This work has made use of the AtomDB Atomic Spectra Database.   }

\facility{\textit{XMM-Newton}, \textit{Suzaku}}

\software{HEAsoft, SAS, XSTAR}

\end{document}